\documentclass[twocolumn,english,aps,pra,superscriptaddress, eqsecnum]{revtex4-1}
\usepackage[T1]{fontenc}
\usepackage[latin9]{inputenc}
\usepackage{xcolor}
\usepackage{amsmath}
\usepackage{amssymb}
\usepackage{graphicx}
\usepackage{esint}
\usepackage{babel}
\begin{document}
\title{Schr\" odinger cats and steady states in subharmonic generation with
Kerr nonlinearities}
\author{Feng-Xiao Sun}
\affiliation{State Key Laboratory for Mesoscopic Physics and Collaborative Innovation
Center of Quantum Matter, School of Physics, Peking University, Beijing
100871, China}
\affiliation{Institute of Theoretical Atomic, Molecular and Optical Physics (ITAMP),
Harvard University, Cambridge, Massachusetts 02138, USA}
\affiliation{Centre for Quantum and Optical Science, Swinburne University of Technology,
Melbourne 3122, Australia}
\affiliation{Collaborative Innovation Center of Extreme Optics, Shanxi University,
Taiyuan, Shanxi 030006, China}
\author{Qiongyi He}
\email{qiongyihe@pku.edu.cn}

\affiliation{State Key Laboratory for Mesoscopic Physics and Collaborative Innovation
Center of Quantum Matter, School of Physics, Peking University, Beijing
100871, China}
\affiliation{Collaborative Innovation Center of Extreme Optics, Shanxi University,
Taiyuan, Shanxi 030006, China}
\affiliation{\textcolor{black}{Beijing Academy of Quantum Information Sciences,
Haidian District, Beijing 100193, China}}
\affiliation{\textcolor{black}{Nano-optoelectronics Frontier Center of the Ministry
of Education, Beijing 100871, China}}
\author{Qihuang Gong}
\affiliation{State Key Laboratory for Mesoscopic Physics and Collaborative Innovation
Center of Quantum Matter, School of Physics, Peking University, Beijing
100871, China}
\affiliation{Collaborative Innovation Center of Extreme Optics, Shanxi University,
Taiyuan, Shanxi 030006, China}
\affiliation{\textcolor{black}{Beijing Academy of Quantum Information Sciences,
Haidian District, Beijing 100193, China}}
\affiliation{\textcolor{black}{Nano-optoelectronics Frontier Center of the Ministry
of Education, Beijing 100871, China}}
\author{Run Yan Teh}
\affiliation{Centre for Quantum and Optical Science, Swinburne University of Technology,
Melbourne 3122, Australia}
\author{Margaret D. Reid}
\affiliation{Institute of Theoretical Atomic, Molecular and Optical Physics (ITAMP),
Harvard University, Cambridge, Massachusetts 02138, USA}
\affiliation{Centre for Quantum and Optical Science, Swinburne University of Technology,
Melbourne 3122, Australia}
\author{Peter D. Drummond}
\email{peterddrummond@gmail.com}

\affiliation{Centre for Quantum and Optical Science, Swinburne University of Technology,
Melbourne 3122, Australia}
\affiliation{Institute of Theoretical Atomic, Molecular and Optical Physics (ITAMP),
Harvard University, Cambridge, Massachusetts 02138, USA}
\affiliation{Weizmann Institute of Science, PO Box 26 Rehovot 7610001, Israel.}
\begin{abstract}
We discuss general properties of the equilibrium state of parametric
down-conversion in superconducting quantum circuits with detunings
and Kerr anharmonicities, in the strongly nonlinear regime. By comparing
moments of the steady state and those of a Schr\" odinger cat, we
show that true Schr\" odinger cats cannot survive in the steady state
if there is any single-photon loss. A delta-function `cat-like' steady-state
distribution can be formed, but this only exists in the limit of an
extremely large nonlinearity. The steady state is a mixed state, which
is more complex than a mixture or linear combination of delta-functions,
and whose purity is reduced by driving. We expect this general behaviour
to occur in other driven, dissipative quantum subharmonic non-equilibrium
open systems.
\end{abstract}
\maketitle

\section{Introduction}

The Schr\" odinger cat is a famous thought experiment~\citep{schrodinger1935gegenwartige},
where a cat is placed in a quantum superposition of two macroscopically
distinct states, either alive or dead. It opens the fundamental question
of whether quantum theory holds true in the macroscopic world~\citep{haroche2013nobel,wineland2013nobel,arndt2014testing}.
Macroscopic superpositions have been experimentally realized in atoms~\citep{monroe1996schrodinger,leibfried2005creation,kovachy2015quantum,omran2019generation}
and photons~\citep{ourjoumtsev2007generation,afek2010high,kirchmair2013observation},
and have been proposed in quantum computation~\citep{mirrahimi2014dynamically},
quantum teleportation~\citep{van2001entangled}, quantum metrology~\citep{joo2011quantum}
and quantum key distribution~\citep{simon2014entangled}. One of
the most common recent strategies for Schr\" odinger cats \citep{reid1992effect}
is via non-equilibrium subharmonic generation \citep{drummond1980non,drummond1981non}
leading to discrete time symmetry-breaking or time crystals \citep{zhang2017observation},
and this approach is analyzed in greater detail here.

The steady state of above-threshold subharmonic generation is known
~for parametric down-conversion without anharmonicities \citep{drummond1980non,wolinsky1988quantum}.
In this case transient Schr\" odinger cats are possible~\citep{reid1992effect,krippner1994transient,munro1995transient}.
Quantum subharmonic generation with Kerr anharmonicities was recently
achieved in superconducting circuits~\citep{leghtas2015confining},
and large cat states were observed. In this experiment, the physics
of the quantum steady state is different from previous studies~\citep{sun2019discrete}.
This exact solution for the steady state demonstrates how dissipation
restores broken time symmetry, with potential applications to solving
combinatorial optimization problems \citep{mcmahon2016fully}.

Quantum optical and quantum circuit physics are similar, except that
quantum circuits operate at microwave instead of optical frequencies.
General driven quantum subharmonic generation with damping and weak
nonlinearities was studied in a previous paper~\citep{sun2019discrete},
where non-equilibrium quantum tunneling \citep{drummond1989quantum}
occurs. Here we focus on the cat-like properties of the steady states
in the case of strong combined parametric and Kerr nonlinearities,
as found in superconducting quantum circuits.

We analytically calculate the exact steady state in subharmonic generation
with strong parametric and Kerr nonlinearities. This exactly soluble
model has a very rich structure, while displaying the expected physics
of more complex devices. We use the resulting exact correlation function
to show that neither simple mixtures of coherent states nor Schr\" odinger
cat states can occur in the steady state. This is confirmed by a numerical
steady-state calculation in the number state basis.

We expect this physical result to occur in other parametric experiments
with a similar dissipative, non-equilibrium behavior. A steady-state
mixture of coherent states~\citep{wolinsky1988quantum} is achievable
as a limiting case of extremely strong nonlinearities, but it is still
a mixed state. This is consistent with the superconducting experiment~\citep{leghtas2015confining}
where an approximate Schr\" odinger cat was observed in a transient
regime. The steady-state in the zero loss case can show a macroscopic
superposition, although it is not unique, due to conserved number
parity.

The outline of this paper is as follows. In Section (\ref{sec:Combined-nonlinearity-model})
we explain our model definitions and notation, with a comparison to
Josephson junction superconducting circuit theory. In Section (\ref{sec:Exact-steady-state-solution})
we obtain the exact steady-state solution, and explain the integration
contour for the complex P-representation manifold. Section (\ref{sec:Numerical-diagonalization})
gives the diagonalization method as another alternative. In Section
(\ref{sec:Moments-and-Schrodinger}), moments are calculated both
exactly and in approximations using coherent or incoherent combinations
of coherent state delta-functions, for comparison purposes. Finally,
Section (\ref{sec:Summary}) summarizes our results.

\section{Combined nonlinearity model\label{sec:Combined-nonlinearity-model}}

\begin{figure}
\centering \includegraphics[width=0.46\textwidth]{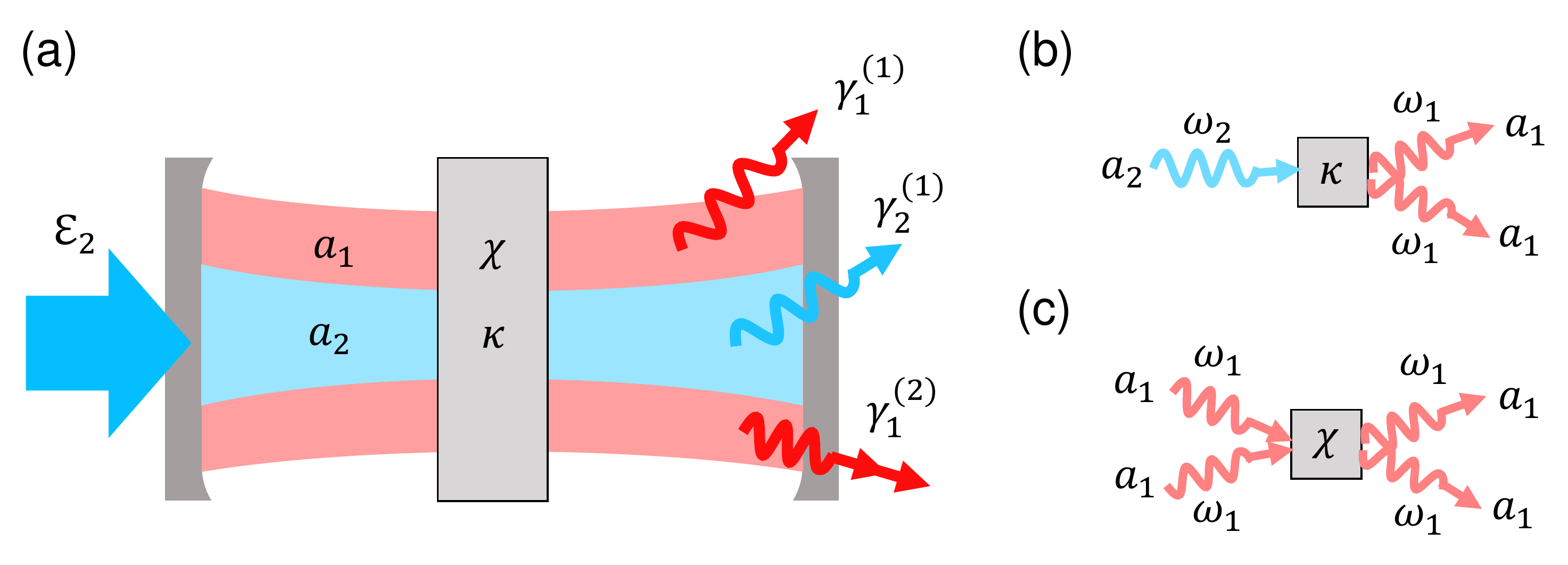}

\caption{(a) Schematic figure of the degenerate parametric oscillator. (b)
Schematic figure of the parametric down-conversion in the system.
(c) Schematic figure of the Kerr nonlinearity for the mode $a_{1}$.}

\label{fig:schematic}
\end{figure}

Firstly, we summarize the system properties and theoretical techniques
used previously \citep{drummond1981non,bartolo2016exact,sun2019discrete},
with both Kerr and parametric nonlinearities. We then treat the detailed
properties of the strongly coupled case. A schematic figure of the
experimental system is shown in Fig.~\ref{fig:schematic}. The annihilation
and creation operators of the $k$-th mode in two coupled resonant
cavities are $a_{k},a_{k}^{\dagger}$ at frequencies $\omega_{k}$.
The frequencies have been set as $\omega_{2}\simeq2\omega_{1}$, so
the system can be externally driven simultaneously at fundamental
and subharmonic frequencies, with $2\omega_{0}$ and $\omega_{0}$,
although we include detunings as well.

\subsection{Hamiltonian}

We assume a doubly resonant nonlinear cavity with a non-interacting
Hamiltonian in the rotating frame of $H_{0}=\hbar\sum\Delta_{k}a_{k}^{\dagger}a_{k}$,
where $\Delta_{k}=\omega_{k}-k\omega_{0}\ll\omega_{0}$ for input
lasers' frequencies of $\omega_{0}$ and $2\omega_{0}$. Then we will
reduce the driving on the subharmonic mode to zero, thus only the
fundamental mode is driven as in the experiment~\citep{leghtas2015confining}.
The interaction Hamiltonian is assumed to be given by
\begin{eqnarray}
H_{I} & = & \hbar\frac{\chi}{2}a_{1}^{\dagger2}a_{1}^{2}+\left(i\hbar\frac{\kappa}{2}a_{2}a_{1}^{\dagger2}+i\hbar\mathcal{E}_{2}a_{2}^{\dagger}+h.c.\right)\,.\label{eq:Hamiltonian}
\end{eqnarray}
Here $\mathcal{E}_{2}$ is the envelope amplitude of the driving for
the mode $a_{2}$, while $\kappa$, $\chi$ are the parametric and
Kerr nonlinearities \citep{drummond1980quantum} respectively. Kerr
nonlinearities are only included for the mode $a_{1}$.

In addition, we include single-photon and two-photon losses in this
open system. Defining $H=H_{0}+H_{I}$, the master equation for the
density matrix $\rho$ is
\begin{equation}
\dot{\rho}=-\frac{i}{\hbar}\left[H,\rho\right]+\sum_{k,j>0}\frac{\gamma_{k}^{(j)}}{j}\mathcal{L}_{k}^{(j)}\left[\rho\right]\,.
\end{equation}
Here $\gamma_{k}^{(j)}$ are the relaxation rates for $j$-photon
losses in the $k$-th mode, with no two-photon losses in mode $k=2$
for simplicity. The dissipative terms are
\begin{align}
\mathcal{L}_{k}^{(j)}\left[\rho\right] & =2\hat{O}\rho\hat{O}^{\dagger}-\rho\hat{O}^{\dagger}\hat{O}-\hat{O}^{\dagger}\hat{O}\rho\,,
\end{align}
where $\hat{O}=\hat{a}_{k}^{j}$. The corresponding thermal noises
are set to zero. This allows us to study the steady-state properties
in the low-temperature limit, in order to understand this exactly
soluble case of maximal quantum coherence.

\subsection{Effective Hamiltonian and master equation}

We suppose the second harmonic mode is strongly damped, as in the
recent Yale experiments \citep{leghtas2015confining}, giving complex
single-photon loss terms defined as $\gamma_{k}=\gamma_{k}^{(1)}+i\Delta_{k}$,
with single-photon losses $\gamma_{k}^{(1)}$ and detunings $\Delta_{k}$
in the $k-$th mode. An adiabatic Hamiltonian is obtained for $a\equiv a_{1}$
as:
\begin{equation}
\frac{H_{A}}{\hbar}=\Delta_{1}a^{\dagger}a+i\left[\frac{\mathcal{E}}{2}a^{\dagger2}-h.c.\right]+\frac{\chi_{e}}{2}a^{\dagger2}a^{2},\label{Hamiltonian-adiabatic}
\end{equation}
 The effective driving field $\epsilon$ and nonlinearity $\chi_{e}$
are: 
\begin{equation}
\mathcal{E}=\frac{\kappa}{\gamma_{2}}\mathcal{E}_{2},\;\chi_{e}=\chi-\frac{\Delta_{2}}{2}\left|\frac{\kappa}{\gamma_{2}}\right|^{2}.
\end{equation}
The master equation of the reduced density matrix $\rho_{1}=\text{Tr}_{2}(\rho)$
is then obtained as 
\begin{align}
\frac{\partial}{\partial t}\rho_{1} & =\frac{1}{i\hbar}[H_{A},\rho_{1}]+\gamma_{1}^{(1)}(2a\rho_{1}a^{\dagger}-a^{\dagger}a\rho_{1}-\rho_{1}a^{\dagger}a)\nonumber \\
 & +\frac{\gamma_{e}^{(2)}}{2}(2a^{2}\rho_{1}a^{\dagger2}-a^{\dagger2}a^{2}\rho_{1}-\rho_{1}a^{\dagger2}a^{2})\,,\label{eq:master_eq}
\end{align}
with an effective two-photon loss $\gamma_{e}^{(2)}$ , where 
\begin{equation}
\gamma_{e}^{(2)}=\gamma_{1}^{(2)}+\frac{\gamma_{2}^{(1)}}{2}\left|\frac{\kappa}{\gamma_{2}}\right|^{2}.\label{eq:parameter}
\end{equation}
Here we have taken the detuning $\Delta_{2}$ into account. Hence
the expression of the effective parameters are slightly different
from those in the previous work~\citep{sun2019discrete}, while the
master equation~(\ref{eq:master_eq}) takes the same general form.

\subsection{Josephson model}

In this subsection, we clarify the relations between the superconducting
Josephson junction experiment~\citep{leghtas2015confining} and our
work. In the supplemental material of the experiment~\citep{leghtas2015confining},
the derivation of the system Hamiltonian, similar to ours~(\ref{eq:Hamiltonian}),
has been provided in detail. Here we will make a brief comparison,
so that we can connect the parameters in our Hamiltonian~(\ref{eq:Hamiltonian})
to those in the experiment~\citep{leghtas2015confining}.

In the experiment~\citep{leghtas2015confining}, two superconducting
microwave oscillators were coupled through a Josephson junction. These
oscillators are the fundamental modes of two superconducting cavities.
One is a high Q cavity termed \textquotedblleft the storage'', where
the steady states formed. The other is a low Q cavity termed \textquotedblleft the
readout\textquotedblright , to evacuate entropy from the storage cavity.
The system Hamiltonian of the qubit, the readout and storage modes
reads
\begin{align}
\frac{H}{\hbar} & =\sum_{m=q,r,s}\omega_{m}n_{m}-\frac{E_{J}}{\hbar}\left(\cos(\varphi)+\frac{\varphi^{2}}{2}\right)\nonumber \\
 & +2\Re(\epsilon_{p}e^{-i\omega_{p}t}+\epsilon_{d}e^{-i\omega_{d}t})(a_{r}^{\dagger}+a_{r}),\nonumber \\
\varphi & =\sum_{m=q,r,s}\varphi_{m}n_{m}.
\end{align}
Here $a_{m}$ is the annihilation operator for the qubit $m=q$, the
readout mode $m=r$ and storage mode $m=s$, respectively, and $n_{m}=a_{m}^{\dagger}a_{m}$
is the corresponding number operator. $E_{J}$ is the Josephson energy,
and $\varphi$ is the phase across the junction, which can be decomposed
as the linear combination of the phase across each mode, with $\varphi_{m}$
denoting the contribution of mode $m$ to the zero point fluctuations
of $\varphi$. The system is irradiated by the drive and pump inputs
with complex amplitudes $\epsilon_{d}$ , $\epsilon_{p}$ and frequencies
$\omega_{d}$, $\omega_{p}$, respectively.

In order to eliminate the system frequencies and the pump amplitude,
we make use of the rotating frame of 
\begin{equation}
U=\exp\left[it\left(\omega_{q}n_{q}+\omega_{d}n_{r}+\frac{\omega_{p}+\omega_{d}}{2}n_{s}\right)-\tilde{\xi}_{p}a_{r}^{\dagger}+\tilde{\xi}_{p}^{*}a_{r}\right],
\end{equation}
 with $\tilde{\xi}_{p}\approx\xi_{p}e^{-i\omega_{p}t}$ and $\xi_{p}\approx-i\epsilon_{p}/\left(\frac{\kappa_{r}}{2}+i\left(\omega_{r}-\omega_{p}\right)\right)$.
Thus the Hamiltonian takes the form of 
\begin{align}
\tilde{H}/\hbar & =\left(\omega_{r}-\omega_{d}\right)n_{r}+\left(\omega_{s}-\frac{\omega_{p}+\omega_{d}}{2}\right)n_{s}\nonumber \\
 & \quad-\frac{E_{J}}{\hbar}\left(\cos(\tilde{\varphi})+\tilde{\varphi}^{2}/2\right),\nonumber \\
\tilde{\varphi} & =\sum_{m=q,r,s}\varphi_{m}\left(\tilde{a}_{m}+\tilde{a}_{m}^{\dagger}\right)+\left(\tilde{\xi}_{p}+\tilde{\xi}_{p}^{*}\right)\varphi_{r},\nonumber \\
\tilde{a}_{q} & =e^{-i\omega_{q}t}a_{q},\tilde{a}_{r}=e^{-i\omega_{d}t}a_{r},\tilde{a}_{s}=e^{-i\frac{\omega_{p}+\omega_{d}}{2}t}a_{s},
\end{align}

If we expand the term $\cos(\tilde{\varphi})$ up to the fourth order,
and only keep non-rotating terms, the Josephson Hamiltonian then reads,
\begin{equation}
\tilde{H}\approx H_{\mathrm{shift}}+H_{\mathrm{Kerr}}+H_{2},
\end{equation}
with 
\begin{align}
\frac{H_{\text{ shift }}}{\hbar} & =\left(-\delta_{q}-\chi_{qr}\left|\xi_{p}\right|^{2}\right)n_{q}\nonumber \\
 & +\left(\omega_{r}-\omega_{d}-\delta_{r}-2\chi_{rr}\left|\xi_{p}\right|^{2}\right)n_{r}\nonumber \\
 & +\left(\omega_{s}-\frac{\omega_{p}+\omega_{d}}{2}-\delta_{s}-\chi_{rs}\left|\xi_{p}\right|^{2}\right)n_{s},\nonumber \\
\frac{H_{\text{ Kerr }}}{\hbar} & =-\sum_{m=q,r,s}\frac{\chi_{mm}}{2}a_{m}^{\dagger}a_{m}^{\dagger}-\chi_{qr}n_{q}n_{r}\nonumber \\
 & \quad-\chi_{qs}n_{q}n_{s}-\chi_{rs}n_{r}n_{s},\nonumber \\
\frac{H_{2}}{\hbar} & =g_{2}^{*}a_{s}^{2}a_{r}^{\dagger}+g_{2}\left(a_{s}^{\dagger}\right)^{2}a_{r}+\epsilon_{d}a_{r}^{\dagger}+\epsilon_{d}^{*}a_{r}.
\end{align}

Here, the Hamiltonian $H_{\text{ Kerr }}$ corresponds to self-Kerr
and cross-Kerr coupling terms, with $\chi_{mm}=\frac{E_{J}}{\hbar}\varphi_{m}^{4}/2$
and $\chi_{mm^{\prime}}=\frac{E_{J}}{\hbar}\varphi_{m}^{2}\varphi_{m^{\prime}}^{2}$.
In the Hamiltonian $H_{2}$, the first two terms are nonlinear couplings
between the storage and readout modes with $g_{2}=\chi_{sr}\xi_{p}^{*}/2$,
which lead to the subharmonic generation. The other terms corresponds
to the weak coherent drive $\epsilon_{d}$ on the readout mode.

\subsection{Josephson parameters}

In this paper we will focus on the evolution of the storage and readout
modes. As given in the supplemental material of Ref.~~\citep{leghtas2015confining},
the Hamiltonian for the reduced system is
\begin{align}
\frac{H_{sr}}{\hbar} & =\Delta_{d}n_{r}+\frac{\Delta_{p}+\Delta_{d}}{2}n_{s}\nonumber \\
 & +g_{2}^{*}a_{s}^{2}a_{r}^{\dagger}+g_{2}\left(a_{s}^{\dagger}\right)^{2}a_{r}+\epsilon_{d}a_{r}^{\dagger}+\epsilon_{d}^{*}a_{r}\nonumber \\
 & -\chi_{rs}n_{r}n_{s}-\sum_{m=r,s}\frac{\chi_{mm}}{2}a_{m}^{\dagger}a_{m}^{2},
\end{align}
where $\Delta_{d}=\omega_{r}-\omega_{d}-\delta_{r}-2\chi_{rr}\left|\xi_{p}\right|^{2}$
and $\Delta_{p}=-\Delta_{d}+2\left(\omega_{s}-\frac{\omega_{p}+\omega_{d}}{2}-\delta_{s}-\chi_{rs}\left|\xi_{p}\right|^{2}\right)$.
In order to include the losses and the quantum noises, master equations
have been analyzed in Ref.~\citep{leghtas2015confining}, where the
single photon damping $\sqrt{\kappa_{r}}a_{r}$ and $\sqrt{\kappa_{s}}a_{s}$
have been considered.

In our notations, we set $a_{1}=a_{s}$ and $a_{2}=a_{r}$. Hence,
this is similar to our initial Hamiltonian~(\ref{eq:Hamiltonian}),
with $\Delta_{1}=(\Delta_{p}+\Delta_{d})/2$, $\Delta_{2}=\Delta_{d}$,
$\mathcal{E}_{2}=-i\epsilon_{d}$, $\kappa=2g_{2}$, $\chi=-\chi_{ss}$,
$\gamma_{1}^{(1)}=\kappa_{s}/2$ and $\gamma_{2}^{(1)}=\kappa_{r}/2$.
In our initial Hamiltonian~(\ref{eq:Hamiltonian}), we have omitted
the cross-Kerr term $\chi_{rs}$ and the self-Kerr term on the second
harmonic mode $\chi_{rr}$ for simplicity.

In fact, the same approximation was used to derive the adiabatic Hamiltonian
in Ref.~\citep{leghtas2015confining} as well. In their supplemental
material, they have shown that the effect of the cross-Kerr term is
negligibly small and thus can be ignored. Since our main results are
obtained under the adiabatic approximation, these omissions are valid
in our situation.

With the detunings and $\chi_{rr}$ omitted, the adiabatic approximation
can be applied in the region where $g_{2}/\kappa_{r},\epsilon_{d}/\kappa_{r},\chi_{rs}/\kappa_{r}\sim\delta$
and $\chi_{ss}/\kappa_{r},\kappa_{s}/\kappa_{r}\sim\delta^{2}$ with
the small dimensionless parameter $\delta\ll1$. By neglecting terms
of order $\delta$ and higher, the adiabatic Hamiltonian has been
derived in the supplemental material of Ref.~\citep{leghtas2015confining},
which reads
\begin{equation}
H_{s}=\epsilon_{2}^{*}a_{s}^{2}+\epsilon_{2}\left(a_{s}^{\dagger}\right)^{2}-\frac{\chi_{ss}}{2}a_{s}^{\dagger^{2}}a_{s}^{2}.
\end{equation}
The corresponding master equation takes the form:
\begin{equation}
\frac{d}{dt}\rho_{s}=-i\left[H_{s},\rho_{s}\right]+\frac{\kappa_{2}}{2}\mathcal{L}\left[a_{s}^{2}\right]\rho_{s}+\frac{\kappa_{s}}{2}\mathcal{L}\left[a_{s}\right]\rho_{s},
\end{equation}
with $\kappa_{2}=4\left|g_{2}\right|^{2}/\kappa_{r}$ and $\epsilon_{2}=-2ig_{2}\epsilon_{d}/\kappa_{r}$.
Compared with our adiabatic Hamiltonian~(\ref{Hamiltonian-adiabatic})
and master equation~(\ref{eq:master_eq}), we find the parameter
mappings for this experiment to be: $\mathcal{E}=2\epsilon_{2}$,
$\chi_{e}=-\chi_{ss}$ and $\gamma_{e}^{(2)}=\kappa_{2}$ with $\gamma_{1}^{(2)}=0$
and $\Delta_{1}=\Delta_{2}=0$.

\section{Exact steady-state solution\label{sec:Exact-steady-state-solution}}

This master equation has an exact analytic solution for the steady-state,
including damping, driving and detunings together with all the nonlinear
couplings. We note that this is neither an energy eigenstate nor a
thermal state, but rather a unique non-equilibrium solution to the
steady-state.

\subsection{Complex P-representation}

To obtain the exact solution, we introduce a generalized P-representation~\citep{drummond1980generalised}
transformation of the single-mode density matrix. If we expand the
reduced quantum density matrix in terms of coherent state projection
operators and a complex P-distribution $P\left(\alpha,\alpha^{+},t\right)$
, one then obtains 
\begin{equation}
\hat{\rho}_{1}=\oiint d\alpha d\alpha^{+}P\left(\alpha,\alpha^{+}\right)\frac{\left|\alpha\right\rangle \left\langle \alpha^{+*}\right|}{\left\langle \alpha^{+*}\right|\left.\alpha\right\rangle }\,,
\end{equation}
 where $\left|\alpha\right\rangle $ is a coherent state and $d\alpha d\alpha^{+}$
is a surface integral measure over a closed surface, so that boundary
terms will vanish on integration by parts. The adiabatic Hamiltonian
results in a single-mode Fokker-Planck equation for $P$, 
\begin{equation}
\frac{\partial P}{\partial t}=\left\{ \frac{\partial}{\partial\alpha}\left[\gamma\alpha-\mathcal{E}\left(\alpha\right)\alpha^{+}\right]+\frac{1}{2}\frac{\partial^{2}}{\partial\alpha^{2}}\mathcal{E}\left(\alpha\right)+h.c.\right\} P,\label{eq:FPE}
\end{equation}
where we define $\gamma\equiv\gamma_{1}=\gamma_{1}^{(1)}+i\Delta_{1}$.
We also introduce an effective complex nonlinear decay of $g=\gamma_{e}^{(2)}+i\chi_{e}$,
and a function $\mathcal{E}\left(\alpha\right)=\mathcal{E}-g\alpha^{2}$.
The notation $h.c.$ indicates hermitian conjugate terms obtained
by the replacement of $\alpha\rightarrow\alpha^{+}$, and the conjugation
of all complex parameters. As in our previous work~\citep{sun2019discrete},
we introduce dimensionless parameters: $\epsilon=\mathcal{E}/g$,
$n=\left|\epsilon\right|$, $c=\gamma/\left(gn\right)$, $\tau=\mathcal{E}t$,
$\beta=\alpha/\sqrt{\epsilon}$, and $e^{i\theta}=g/|g|=n/\epsilon$,
so that the Fokker-Planck equation can be simplified to the form:
\begin{eqnarray}
{\normalcolor \frac{\partial P\left(\vec{\beta}\right)}{\partial\tau}} & {\normalcolor =} & {\normalcolor e^{i\theta}\left\{ \frac{\partial}{\partial\beta}\left[c\beta-\left(1-\beta^{2}\right)\beta^{+}\right]+\right.}\nonumber \\
{\normalcolor } & {\normalcolor } & {\normalcolor +\left.\frac{1}{2n}\frac{\partial^{2}}{\partial\beta^{2}}\left(1-\beta^{2}\right)+h.c.\right\} P\left(\vec{\beta}\right).}\label{FPE_scaled}
\end{eqnarray}
With this transformation, time is scaled relative to the two-photon
driving rate. Here $c$ is a complex dimensionless single-photon loss
and detuning, and $n$ is the photon number at which saturation of
the mode occupation occurs due to the nonlinear losses.

The steady-state solution of the scaled Fokker-Planck equation~(\ref{FPE_scaled})
can be derived via the potential method~\citep{graham1971fluctuations,graham1971generalized,risken1972solutions,seybold1974theory}
\begin{equation}
P_{1}\left(\vec{\beta}\right)=N\exp\left[-\Phi\left(\vec{\beta}\right)\right],\label{steady-state-sp}
\end{equation}
where $N$ is a normalization constant and $\Phi$ satisfies 
\begin{eqnarray}
\frac{\left(1-\beta^{2}\right)}{2n}\frac{\partial\Phi}{\partial\beta} & = & (c-\frac{1}{n})\beta-\left(1-\beta^{2}\right)\beta^{+},\nonumber \\
\frac{\left(1-\beta^{+2}\right)}{2n}\frac{\partial\Phi}{\partial\beta^{+}} & = & (c^{*}-\frac{1}{n})\beta^{+}-\left(1-\beta^{+2}\right)\beta.\label{eq:for-potential-sp}
\end{eqnarray}
These equations~(\ref{eq:for-potential-sp}) are obtained by inserting
the form~(\ref{steady-state-sp}) into the Fokker-Planck equation~(\ref{FPE_scaled})
and requiring that $\partial P_{1}/\partial\tau=0$ in the steady-state.

By solving the differential equations~(\ref{eq:for-potential-sp})
directly, the exact steady-state solution with quantum noise can be
expressed via the potential:
\begin{equation}
\Phi\left(\vec{\beta}\right)=-n\left[\beta^{+}\beta+\tilde{c}\ln(1-\beta^{2})+h.c.\right],\label{potential-simple}
\end{equation}
with $\tilde{c}=c-1/n$. Thus, the steady-state probability distribution
is 
\begin{equation}
{\normalcolor P_{S}\left(\vec{\beta}\right)=N\left[(1-\beta^{2})^{\tilde{c}}(1-\beta^{+2})^{\tilde{c}^{\ast}}\exp(2\beta^{+}\beta)\right]^{n}.}\label{probability}
\end{equation}

This is the exact zero-temperature steady-state solution for the density
matrix. Written in this way, we can see how it scales with the effective
driving field $n$ occurring in the exponent. Apart from $n$, all
the parameters here can have complex values, which is necessary when
treating the situations in recent quantum circuit experiments~\citep{leghtas2015confining}.

\begin{figure}
\centering \includegraphics[width=0.35\textwidth]{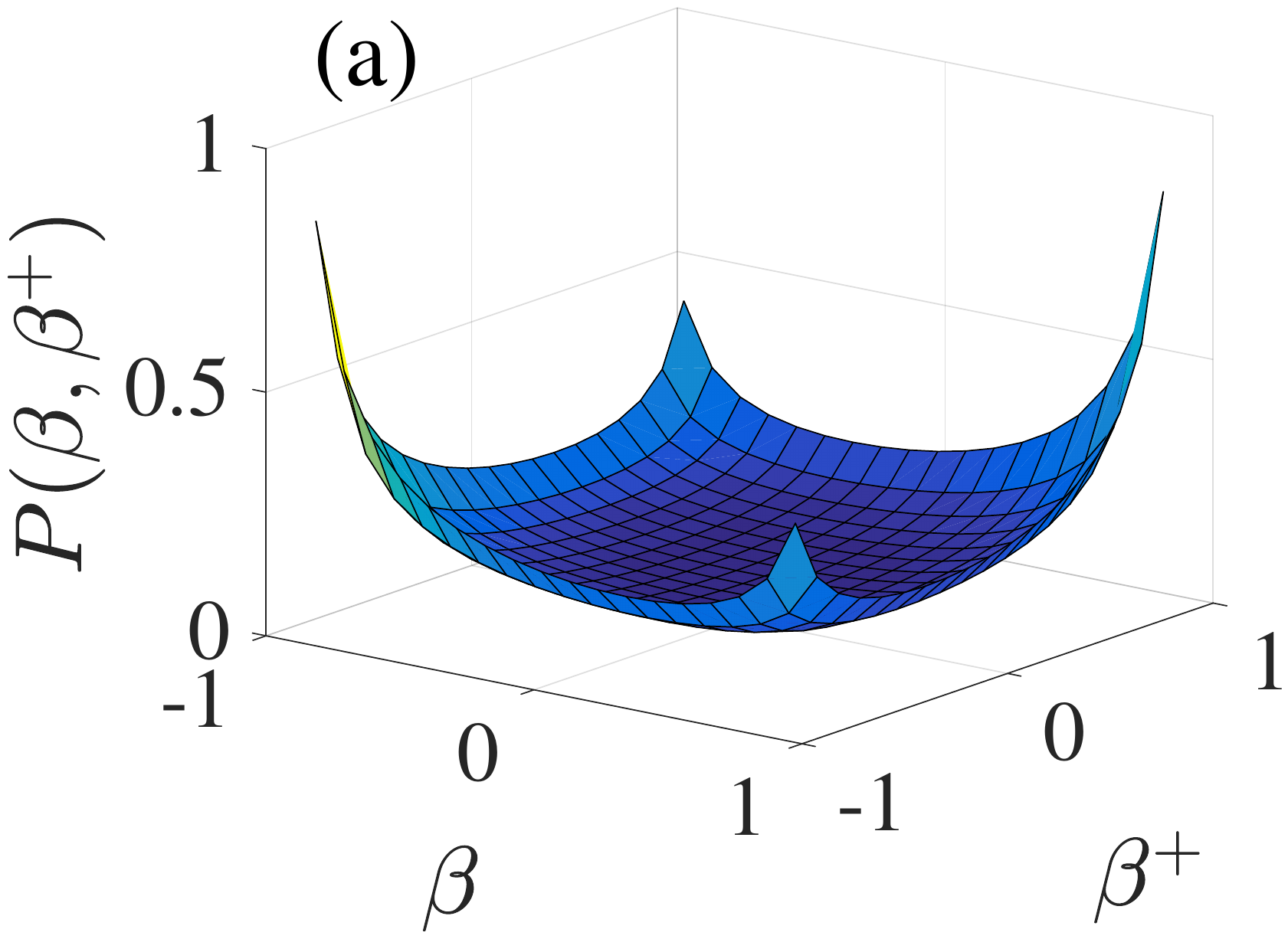}

$\vphantom{}$

\includegraphics[width=0.35\textwidth]{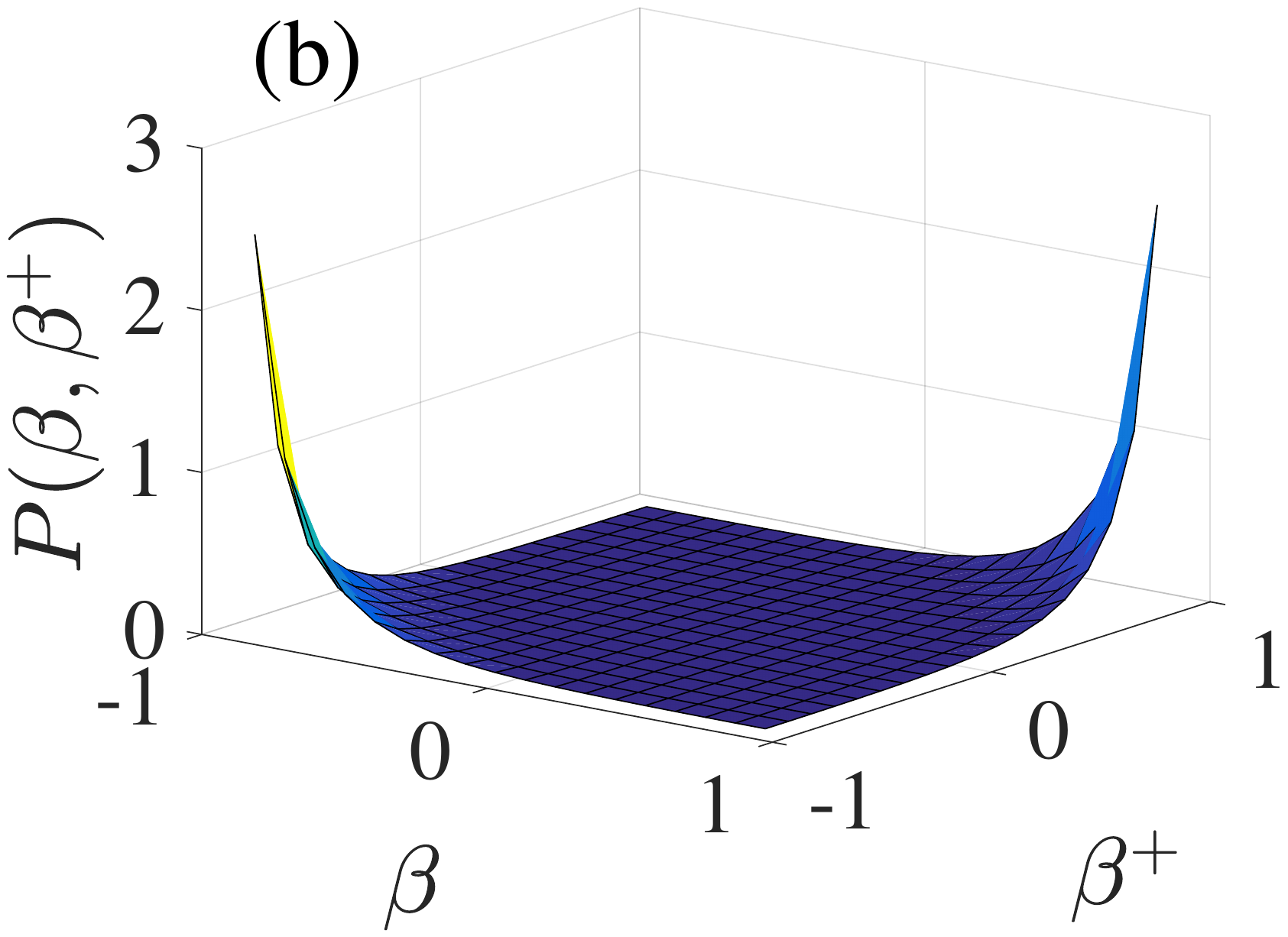}

$\vphantom{}$

\includegraphics[width=0.35\textwidth]{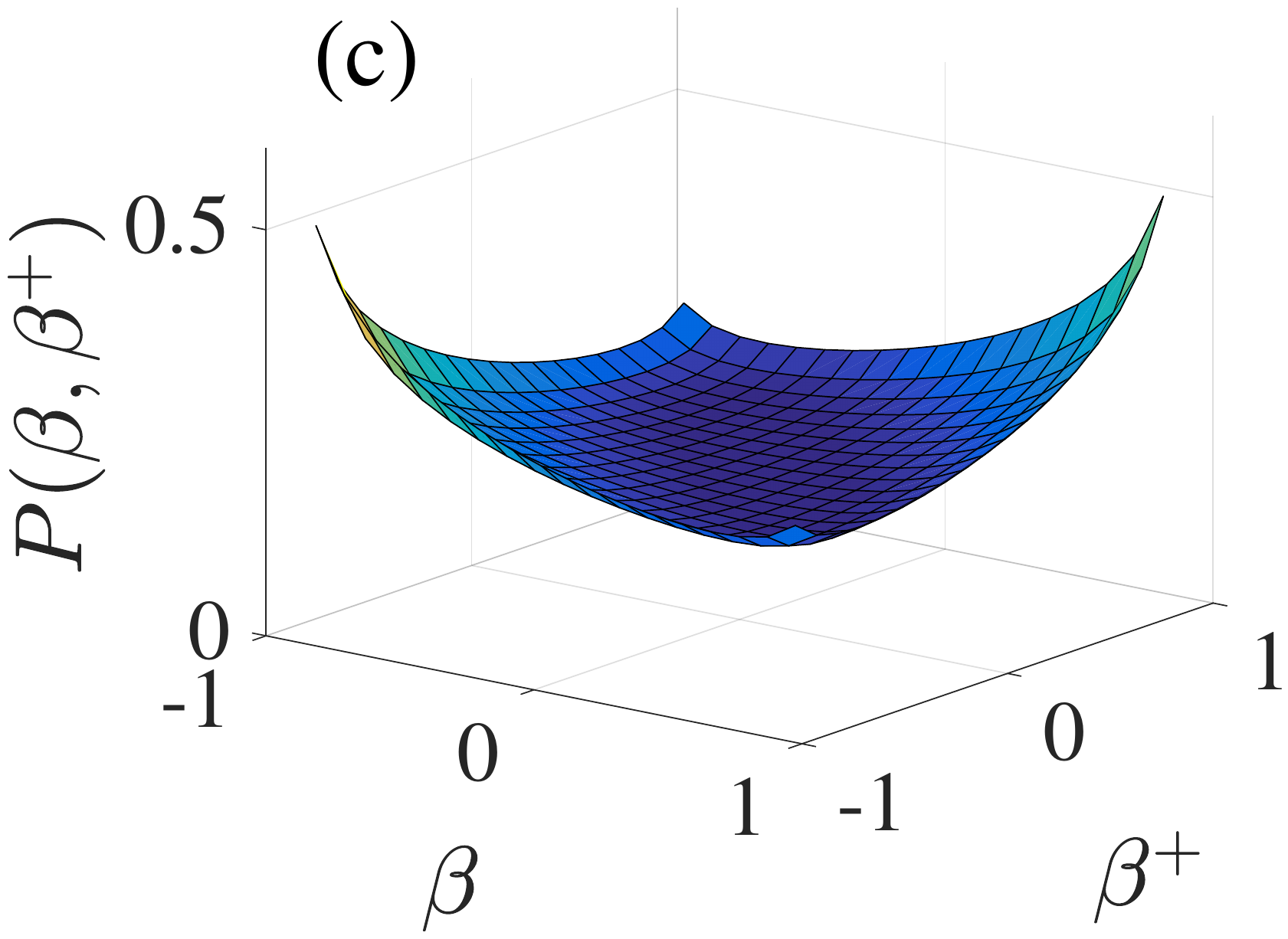}

$\vphantom{}$

\includegraphics[width=0.35\textwidth]{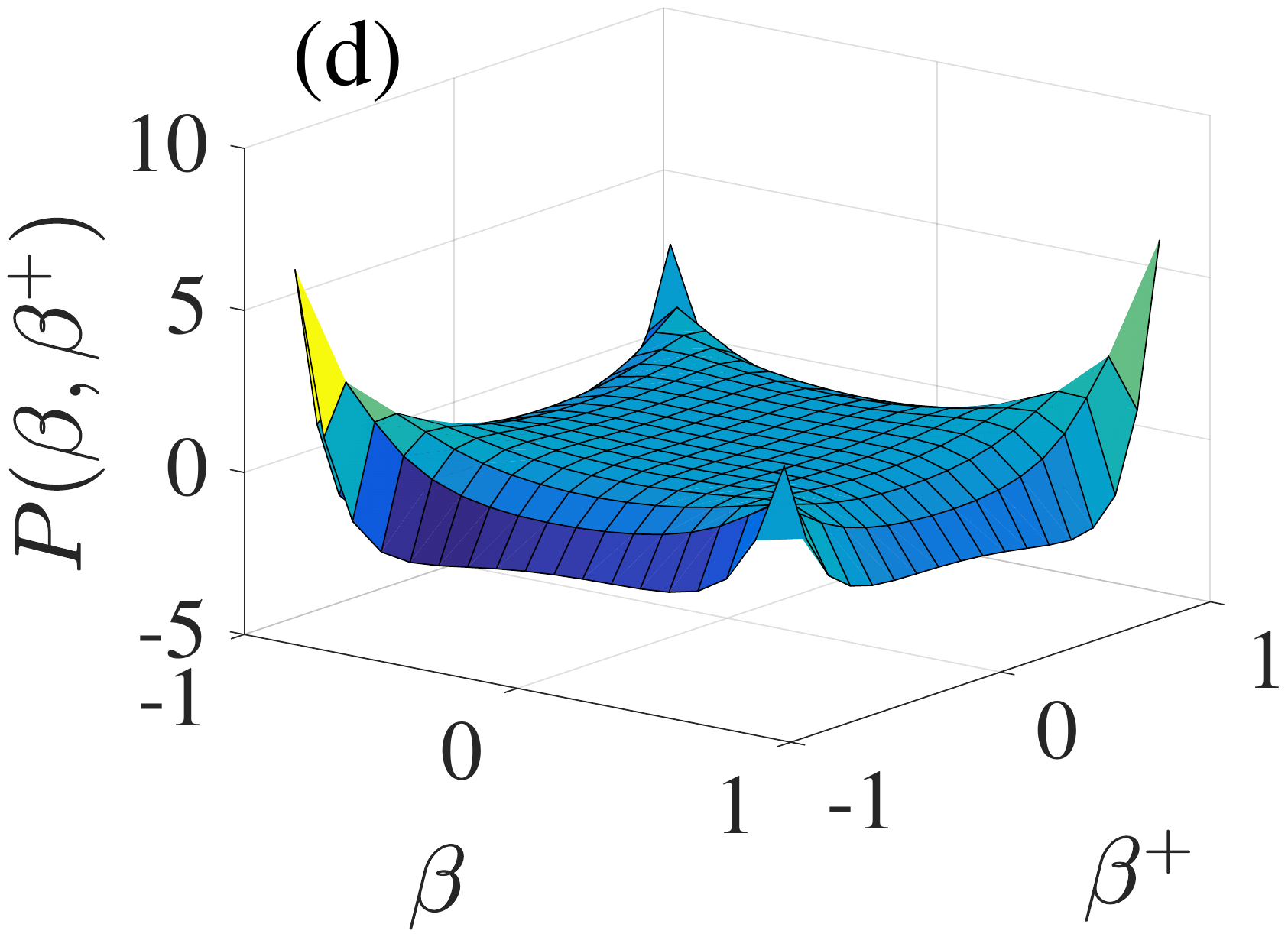}

\caption{Real parts of steady-state probability distributions~(\ref{probability})
for (a) $\tilde{c}=-2.79+0.93i$ and $\epsilon=-0.192-0.097i$, (b)
large $n=\left|\epsilon\right|$: $\tilde{c}=-0.279+0.093i$ and $\epsilon=-1.92-0.97i$,
(c) small $\left|\Re(\tilde{c})\right|$: $\tilde{c}=-0.93+0.93i$
and $\epsilon=-0.192-0.097i$, (d) large $\left|\Im(\tilde{c})\right|$:
$\tilde{c}=-2.79+9.3i$ and $\epsilon=-0.192-0.097i$.\textcolor{brown}{}}

\label{fig:probability}
\end{figure}

In the case where the power $\tilde{c}$ has a negative real part,
if a real planar complex manifold is chosen, one obtains singular
peaks at the boundaries where $\left|\beta\right|,\left|\beta^{+}\right|=\pm1$,
as shown in Fig (\ref{fig:probability}). This would give boundary
terms on partial integration, causing errors. Integration over phase-space
distributions requires vanishing boundary terms. Instead, one must
choose a curved topological structure with cuts on the complex integration
manifold. This leads to branch points, rather than local potential
minima. This is why there is no quantum tunneling, although transient
Schr\" odinger cats can be formed in this type of experiment~\citep{leghtas2015confining}.

As a result, this physical situation requires a completely different
phase-space manifold to that investigated in the previous work~\citep{sun2019discrete},
where the real part of $\tilde{c}$ is positive. In that case, there
is quantum tunneling between local potential minima on a finite, bounded
manifold. \textcolor{blue}{} To define the distribution for strong
coupling, one must choose complex integration contours which are closed,
continuous \citep{drummond1980generalised,drummond1981non,bartolo2016exact},
and without boundaries. This is obtained by inserting cuts at the
branch-points for $\beta=\pm1$ and $\beta^{+}=\pm1$, combined with
complex Pochhammer contours. This method is used to represent the
beta and hypergeometric special functions \citep{pochhammer1890theorie,macrobert1938functions,bateman1955higher}.
One way to visualize this is to imagine the contours drawn on both
sides of two sheets of paper, one for $\beta$ and one for $\beta^{+}$.

\subsection{Moments and Correlations}

The second-order correlation function of the single-mode intra-cavity
field is defined as\textcolor{blue}{}
\begin{equation}
g^{(2)}(0)=\frac{\langle a^{\dagger}a^{\dagger}aa\rangle}{\langle a^{\dagger}a\rangle^{2}},
\end{equation}
where the $k$-th moment can be calculated with P-representation integrals
as
\begin{equation}
I_{kk'}=\langle a^{\dagger k}a^{k'}\rangle=\oiint(\epsilon^{*})^{\frac{k}{2}}\epsilon^{\frac{k'}{2}}\beta^{+k}\beta^{k'}P_{S}(\beta,\beta^{+})d\beta^{+}d\beta.\label{moment}
\end{equation}
It is well known that nonclassical effects like photon anti-bunching
will occur if $g^{(2)}(0)<1$ and classical bunching takes place if
$g^{(2)}(0)>1$. Thus, $g^{(2)}(0)$ is often used to distinguish
classical from non-classical behavior ~\citep{scully1997quantum}.

The exact solution for the moments~\citep{drummond1981non} is obtained
by expanding the term $e^{2n\beta^{+}\beta}=\sum_{m}\left(2n\right)^{m}\beta^{m}\beta^{+m}/m!$
in Eq.~(\ref{probability}). In this way, we obtain the form of moment
after normalization and integration over the complex manifold, as:
\begin{eqnarray}
I_{kk'}^{ex} & = & N'\sum_{m}\frac{(2n)^{m}}{m!}(-\sqrt{\epsilon})^{k'}(-\sqrt{\epsilon^{*}})^{k}\nonumber \\
 &  & \times{}_{2}F_{1}(-m-k',n\tilde{c}+1,2n\tilde{c}+2,2)\nonumber \\
 &  & \times{}_{2}F_{1}(-m-k,n\tilde{c}^{\ast}+1,2n\tilde{c}^{\ast}+2,2)\,.\label{eq:moment-expand}
\end{eqnarray}
Here $_{2}F_{1}$ is the hypergeometric function, and $N'$ is the
normalization factor,
\begin{eqnarray}
N'^{-1} & = & \sum_{m}\frac{(2n)^{m}}{m!}{}_{2}F_{1}(-m,n\tilde{c}+1,2n\tilde{c}+2,2)\nonumber \\
 &  & \times{}_{2}F_{1}(-m,n\tilde{c}^{\ast}+1,2n\tilde{c}^{\ast}+2,2).
\end{eqnarray}

The case of real $\tilde{c}$ has been investigated in Ref~\citep{wolinsky1988quantum,krippner1994transient},
where there was no anharmonic nonlinearity, and a real manifold was
used. It was suggested that the steady-state distribution approaches
a set of $\delta$ functions in strong coupling limits. The case without
single-photon loss and anharmonic nonlinearity has also been studied
in Ref.~\citep{gilles1994generation}, where one always has $\tilde{c}=-1/n$.
In this case, steady-state Schr\" odinger cats can be achieved with
initial Fock states. Other work studying this potential in different
parameter regimes was used to benchmark our numerical results, given
below \citep{bartolo2016exact}.

\section{Numerical diagonalization\label{sec:Numerical-diagonalization}}

As a comparison and independent check of these exact results, we have
also solved the master equation Eq. (\ref{eq:master_eq}) numerically
by expanding the density operator in a number state basis. The steady
state of the system corresponds to the eigenstate of the Liouvillian
operator with zero eigenvalue. This steady state density operator
is then used to compute the statistical moments of interest. In this
approach, which is valid for small photon number, we numerically diagonalize
the Liouville operator of the master equation, with a photon number
cutoff. This allows us to compare the analytical and numerical approaches.
We find that there is excellent agreement between the two methods.

\subsection{Number state basis}

In order to verify our analytic results, we applied these numerical
number state methods to the same case. We expand the density operator
$\rho$ in the number-state basis, where its matrix elements $\rho_{kl}$
are defined as
\begin{equation}
\rho_{kl}=\langle k|\rho|l\rangle.
\end{equation}
Then the master equation~(\ref{eq:master_eq}) takes the form: 
\begin{equation}
\frac{d}{dt}\rho_{ij}=T_{ij}^{kl}\rho_{kl}.
\end{equation}
Here the Einstein summation convention has been used on identical
indices. And $T_{ij}^{kl}$ is a four-dimensional transition matrix,
which describes the transition from the state $\rho_{kl}$ to the
state $\rho_{ij}$. It can be written as 
\begin{eqnarray}
T_{ij}^{kl} & = & \frac{\mathcal{E}}{2}\sqrt{i(i-1)}\delta_{i;j}^{k+2;l}-\frac{\mathcal{E}}{2}\sqrt{(j+1)(j+2)}\delta_{i;j}^{k;l-2}\nonumber \\
 &  & +\frac{\mathcal{E}^{*}}{2}\sqrt{j(j-1)}\delta_{i;j}^{k;l+2}-\frac{\mathcal{E}^{*}}{2}\sqrt{(i+1)(i+2)}\delta_{i;j}^{k-2;l}\nonumber \\
 &  & -\left[\gamma i+\gamma^{\ast}j+\frac{g}{2}i(i-1)+\frac{g^{\ast}}{2}j(j-1)\right]\delta_{i;j}^{k;l}\nonumber \\
 &  & +\gamma_{e}^{(2)}\sqrt{(i+1)(i+2)(j+1)(j+2)}\delta_{i;j}^{k-2;l-2}\nonumber \\
 &  & +2\gamma_{1}^{(1)}\sqrt{(i+1)(j+1)}\delta_{i;j}^{k-1;l-1},
\end{eqnarray}
with 
\begin{equation}
\delta_{i;j}^{k;l}=\left\{ \begin{array}{ll}
1 & \text{if \ensuremath{i=k} and \ensuremath{j=l},}\\
0 & \text{otherwise.}
\end{array}\right.
\end{equation}
The system can be characterized by the eigenvectors of the transition
matrix $T_{ij}^{kl}$. The steady state of the system corresponds
to the eigenvector with zero eigenvalue~\citep{kinsler1991quantum}.

\subsection{Transition matrix elements}

Within the numerical calculation, we must use a photon number cutoff
$N$ to make the transition matrix finite, $0\le i,j,k,l\le N$. This
approximation is valid if the high-photon-number states play negligible
roles in determining the system's evolution. We check that the cut-off
is set to a high enough value by repeating the calculation with a
higher cutoff and checking that no change occurs.

Hence, the four-dimensional matrix $T_{ij}^{kl}$ can be reduced to
a two-dimensional one $T_{\bar{\alpha}}^{\bar{\beta}}$ with this
truncation, so that 
\begin{equation}
\frac{d}{dt}\rho_{\bar{\alpha}}=T_{\bar{\alpha}}^{\bar{\beta}}\rho_{\bar{\beta}},
\end{equation}
with 
\begin{eqnarray}
T_{\bar{\alpha}}^{\bar{\beta}} & = & \frac{\mathcal{E}}{2}\sqrt{i(i-1)}\delta_{\bar{\alpha}}^{\bar{\beta}+2N+2}-\frac{\mathcal{E}}{2}\sqrt{(j+1)(j+2)}\delta_{\bar{\alpha}}^{\bar{\beta}-2}\nonumber \\
 &  & +\frac{\mathcal{E}^{*}}{2}\sqrt{j(j-1)}\delta_{\bar{\alpha}}^{\bar{\beta}+2}-\frac{\mathcal{E}^{*}}{2}\sqrt{(i+1)(i+2)}\delta_{\bar{\alpha}}^{\bar{\beta}-2N-2}\nonumber \\
 &  & -\left[\gamma i+\gamma^{\ast}j+\frac{g}{2}i(i-1)+\frac{g^{\ast}}{2}j(j-1)\right]\delta_{\bar{\alpha}}^{\bar{\beta}}\nonumber \\
 &  & +\gamma_{e}^{(2)}\sqrt{(i+1)(i+2)(j+1)(j+2)}\delta_{\bar{\alpha}}^{\bar{\beta}-2N-4}\nonumber \\
 &  & +2\gamma_{1}^{(1)}\sqrt{(i+1)(j+1)}\delta_{\bar{\alpha}}^{\bar{\beta}-N-2},\label{eq:transition-matrix-sp}
\end{eqnarray}
where 
\begin{eqnarray}
\bar{\alpha} & = & (N+1)i+j+1,\quad\bar{\beta}=(N+1)k+l+1.
\end{eqnarray}
Here $\delta_{\bar{\alpha}}^{\bar{\beta}}$ is a Kronecker delta,
and $\bar{\alpha}$, $\bar{\beta}$ are in the range of $[1,(N+1)^{2}]$.

We label the $k$-th eigenvalue by $\epsilon_{k}$ and its corresponding
eigenvector by $\rho_{\bar{\alpha}}^{(k)}$ so that 
\begin{equation}
\rho_{\bar{\alpha}}(t)=\sum_{k\ge0}A_{k}\exp(\epsilon_{k}t)\rho_{\bar{\alpha}}^{(k)},
\end{equation}
where the coefficients $A_{k}$ define the initial state. We order
the indices $k$ by the size of the real part of the eigenvalues,
$\Re(\epsilon_{k})\ge\Re(\epsilon_{k+1})$. Therefore, $\epsilon_{0}$
is the stable eigenvalue with $\epsilon_{0}=0$, and $\rho_{\bar{\alpha}}^{(0)}$
corresponds to the stable state. 

With the numerical expansion method, the stable state $\rho_{\bar{\alpha}}^{(0)}$
can be obtained by solving the eigenvalue problem of the transition
matrix $T_{\bar{\alpha}}^{\bar{\beta}}$. Then the average photon
numbers $\left\langle a^{\dagger}a\right\rangle $ and the second
order correlation functions $g^{(2)}(0)$ can be obtained directly
by
\begin{align}
\langle a^{\dagger}a\rangle & =\text{Tr}\left[a^{\dagger}a\rho^{(0)}\right],\nonumber \\
g^{(2)}(0) & =\frac{\text{Tr}\left[a^{\dagger}a^{\dagger}aa\rho^{(0)}\right]}{\text{Tr}\left[a^{\dagger}a\rho^{(0)}\right]^{2}},
\end{align}
where $\rho^{(0)}$ is the stable-state matrix reshaped from the stable-state
vector $\rho_{\bar{\alpha}}^{(0)}$. The numerical results are shown
below in Fig.~\ref{fig:compare-c} and \ref{fig:compare-lambda}
below with green dots. They agree with the analytic results very well.
This confirms the validity of our analytic calculations.

\section{\label{sec:Moments-and-Schrodinger}Moments and Schr\" odinger Cat
comparisons}

We will use these exact analytic and approximate numerical results
to check the validity of approximate delta-function steady-state distributions
which we introduce below~(\ref{distribution}). These correspond
to the physical assumptions that one has either a quantum superposition
or a quantum mixture of two coherent states with opposite signs. As
we show below, neither assumption is correct in the steady-state of
this driven, non-equilibrium quantum system.

\subsection{Experimental parameter values}

For numerical evaluations of the steady-state moments, we obtain the
parameters of the recent experiment~\citep{leghtas2015confining},
using the results of Section (\ref{sec:Combined-nonlinearity-model}).
In our notation, we obtain that for these recent quantum circuit experiments,
$\gamma/2\pi=3.98$kHz, $g/2\pi=(7.96-4i)$ kHz and $\mathcal{E}=(-19.2-0.07i)$
kHz. Thus, we have $\tilde{c}=-0.279+0.093i$ and $\epsilon=-1.92-0.97i$.
Since the real part of $\tilde{c}$ is negative, there will be singularities
occurring at $\beta=\pm1$ or $\beta^{+}=\pm1$.

From now on, we will treat the strong coupling regime, which corresponds
to the parameter region of $\Re(\tilde{c})<0$. Using the definitions
of $\tilde{c}$ and $g$, we have
\begin{equation}
n\tilde{c}=\frac{\left(\gamma_{1}^{(1)}+i\Delta_{1}\right)\left(\gamma_{e}^{(2)}-i\chi_{e}\right)}{\left(\gamma_{e}^{(2)}\right)^{2}+\chi_{e}^{2}}-1.\label{eq:c}
\end{equation}
Considering that $n>0$, it follows that $\Re(\tilde{c})<0$ is equivalent
to $\gamma_{e}^{(2)}(\gamma_{1}^{(1)}-\gamma_{e}^{(2)})+\chi_{e}(\Delta_{1}-\chi_{e})<0$.
This is satisfied if there is either a weak single-photon damping
$\gamma_{1}^{(1)}$ or strong nonlinear couplings $\chi_{e}$, $\gamma_{e}^{(2)}$.
It is easily checked, provided there are no detunings, that $\Re(\tilde{c})\ge-1/n$
and the limit $\tilde{c}\to-1/n$ occurs if $\gamma_{1}^{(1)}\ll\gamma_{e}^{(2)}$
or $\gamma_{1}^{(1)}\ll\chi_{e}$.

Considering that nonlinear losses are always weak, the relation $\gamma_{1}^{(1)}\ll\gamma_{e}^{(2)}$
can occur with large $\kappa$ refer to Eq.~(\ref{eq:parameter}).
Thus the limit $\tilde{c}\to-1/n$ occurs either with large nonlinearities
$\kappa$ or $\chi$.

\subsection{Delta-function approximations}

To understand the physics more clearly, we note that in the limit
of $\tilde{c}\rightarrow-1/n$, the exact solution is a product of
simple poles with opposite contour integration directions. These can
be integrated using Cauchy's theorem, and correspond to a delta-function
solution, so the ratio of the probabilities at the singularities
is
\begin{equation}
\frac{P_{lim}(\beta=\pm1,\beta^{+}=\pm1)}{P_{lim}(\beta=\pm\sqrt{\lambda_{c}},\beta^{+}=\mp1)}=e^{4n}.
\end{equation}
If we assume this is also true approximately for $\tilde{c}\neq-1/n$,
we obtain a real distribution~\citep{wolinsky1988quantum} in the
form of 
\begin{eqnarray}
P_{lim}(\beta,\beta^{+}) & = & \frac{\delta(\beta-1)\delta(\beta^{+}-1)+\delta(\beta+1)\delta(\beta^{+}+1)}{2(1+e^{-4n})}\nonumber \\
 &  & +\frac{\delta(\beta-1)\delta(\beta^{+}+1)+\delta(\beta+1)\delta(\beta^{+}-1)}{2(1+e^{4n})}.\nonumber \\
\label{distribution}
\end{eqnarray}

We now contrast this with an idealized, even cat state $\left|\psi\right\rangle _{cat}\varpropto\left[\left|\sqrt{\epsilon}\right\rangle +\left|-\sqrt{\epsilon}\right\rangle \right]$,
where the P-representation takes the form after normalization 
\begin{eqnarray}
P_{cat}(\beta,\beta^{+}) & = & \frac{\delta(\beta-1)\delta(\beta^{+}-1)+\delta(\beta+1)\delta(\beta^{+}+1)}{2(1+e^{-2n})}\nonumber \\
 &  & +\frac{\delta(\beta-1)\delta(\beta^{+}+1)+\delta(\beta+1)\delta(\beta^{+}-1)}{2(1+e^{2n})}.\nonumber \\
\label{distribution-cat}
\end{eqnarray}

The factor is $e^{-2n}$ ($e^{2n}$), rather than $e^{-4n}$ ($e^{4n}$)
in Eq.~(\ref{distribution}), so even if the steady state does evolve
to a delta-function distribution~(\ref{distribution}), it will be
a mixed state instead of a true cat state.

In this case, the density matrix can be derived to have the following
form, 
\begin{equation}
\rho_{lim}=p\left|\psi\right\rangle _{cat}\left\langle \psi\right|_{cat}+\left(1-p\right)\rho_{mix}.\label{eq:mixed-cat}
\end{equation}
Here 
\begin{align}
p & =(1+e^{2n})/(1+e^{4n}),\nonumber \\
\rho_{mix} & =\frac{1}{2}\left[\left|\sqrt{\epsilon}\right\rangle \left\langle \sqrt{\epsilon}\right|+\left|-\sqrt{\epsilon}\right\rangle \left\langle -\sqrt{\epsilon}\right|\right].
\end{align}
 The purity of this limiting form can then be obtained as 
\begin{equation}
\mu=Tr\left[\rho_{lim}^{2}\right]=\frac{e^{8n}+6e^{4n}+1}{2(e^{4n}+1)^{2}}\,,
\end{equation}
which is a monotonic decreasing function of $n$ since 
\begin{equation}
\frac{d\mu}{dn}=-\frac{8e^{4n}(e^{4n}-1)}{(e^{4n}+1)^{3}}<0\,,
\end{equation}
 for $n>0$. Thus, the driving will weaken the purity of the steady
state since $n$ is proportional to the driving $\mathcal{E}_{2}$.

It is obvious that we will have $p\to1$ in the limit of $n\to0$.
Thus the delta-function distribution tends to be a true Schr\" odinger
cat state in this limit. However, since $|\epsilon|=n\to0$, the steady
state will actually reduce to a vacuum state. This is natural that
a non-driven system can be expressed as a vacuum state. In the opposite
limit of $n\to\infty$, the delta-function steady-state distribution~(\ref{distribution})
will reduce to the mixed state $\rho_{mix}$ since $p\to0$. Therefore,
a pure Schr\" odinger cat state is unreachable in the steady state
of the system, even using an approximate delta-function solution.

The parity $\hat{\mathcal{P}}=(-1)^{a^{\dagger}a}$ can also be studied
directly with the complex P-distribution~(\ref{distribution}). In
the P-representation, the parity operator is equivalent to the average
of $\mathcal{P}=\exp(-2n\beta^{+}\beta)$. In the steady state of
the delta-function approximation, we have $\mathcal{P}_{ss}=\text{sech}(2n)$.
This means that $\mathcal{P}_{ss}=1$ in the case of $n=0$, and $\mathcal{P}_{ss}=0$
in the limit of $n\to\infty$. It is consistent with the density matrix~(\ref{eq:mixed-cat})
which is a vacuum state when $n=0$ and a mixed state when $n\to\infty$.
Parity is not conserved because of the finite single-photon loss.

\subsection{Steady-state distributions}

The exact steady-state distributions~(\ref{probability}) with different
parameters are shown in Fig.~\ref{fig:probability}, plotted on a
finite manifold. We see that delta-function distribution will be obtained
approximately with large $|\Re(\tilde{c})|$ and small $|\Im(\tilde{c})|$,
and reduced to classical mixture of coherent states with large $n$.
However, these graphs also demonstrate that the probability does not
vanish at the boundaries, which means that with $\Re\left(\tilde{c}\right)<0$
on this bounded manifold, the potential solution when restricted to
this planar manifold is no longer a solution to the original master
equation, since boundary terms from integration by parts are non-vanishing.

An inspection of Fig \ref{fig:probability} shows that when assuming
a real, bounded manifold, the distribution is neither a true delta
function, nor does it vanish at the boundaries, which is the reason
why the exact complex contour manifold is essential when there are
poles.

\subsection{Moment comparisons}

As a result, the true steady states are clearly neither mixtures of
delta functions nor Schr\" odinger cats. This difference can be quantified
by using the steady-state distribution~(\ref{distribution}), to
compare moments. The approximate $k$-th moment is obtained directly
with the definition~(\ref{moment}) as,
\begin{eqnarray}
I_{kk'}^{lim} & = & \frac{(\sqrt{\epsilon})^{k'}(\sqrt{\epsilon^{*}})^{k}+(-\sqrt{\epsilon})^{k'}(-\sqrt{\epsilon^{*}})^{k}}{2\left(1+e^{-4n}\right)}\nonumber \\
 &  & +\frac{(-\sqrt{\epsilon})^{k'}(\sqrt{\epsilon^{*}})^{k}+(\sqrt{\epsilon})^{k'}(-\sqrt{\epsilon^{*}})^{k}}{2\left(1+e^{4n}\right)}.\label{eq:moment-delta}
\end{eqnarray}
Similarly, the moment can be written down directly with the cat state~(\ref{distribution-cat})
as:
\begin{eqnarray}
I_{kk'}^{cat} & = & \frac{(\sqrt{\epsilon})^{k'}(\sqrt{\epsilon^{*}})^{k}+(-\sqrt{\epsilon})^{k'}(-\sqrt{\epsilon^{*}})^{k}}{2\left(1+e^{-2n}\right)}\nonumber \\
 &  & +\frac{(-\sqrt{\epsilon})^{k'}(\sqrt{\epsilon^{*}})^{k}+(\sqrt{\epsilon})^{k'}(-\sqrt{\epsilon^{*}})^{k}}{2\left(1+e^{2n}\right)}\,.\label{eq:moment-cat}
\end{eqnarray}

\begin{figure}
\centering \includegraphics[width=0.23\textwidth]{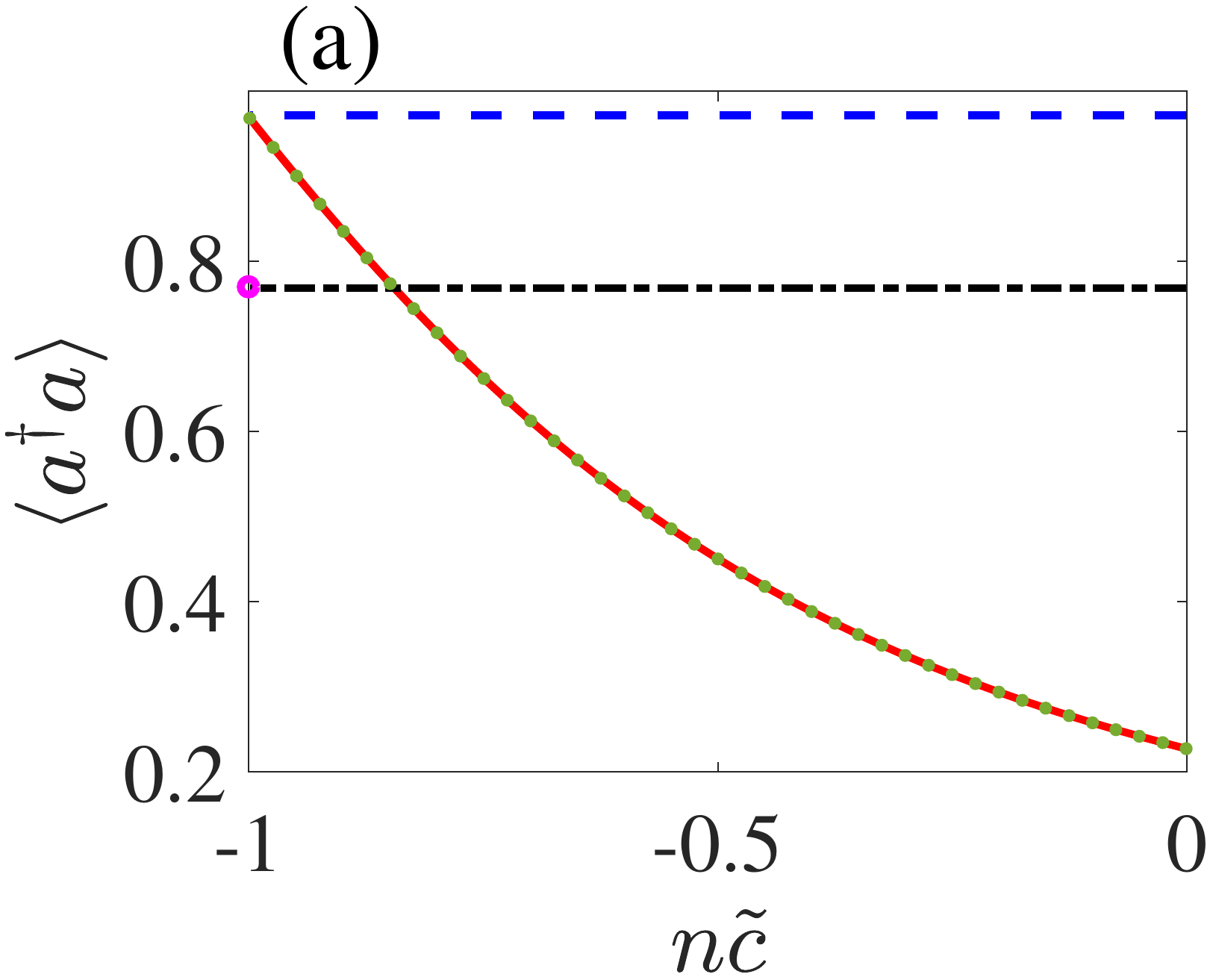} \includegraphics[width=0.23\textwidth]{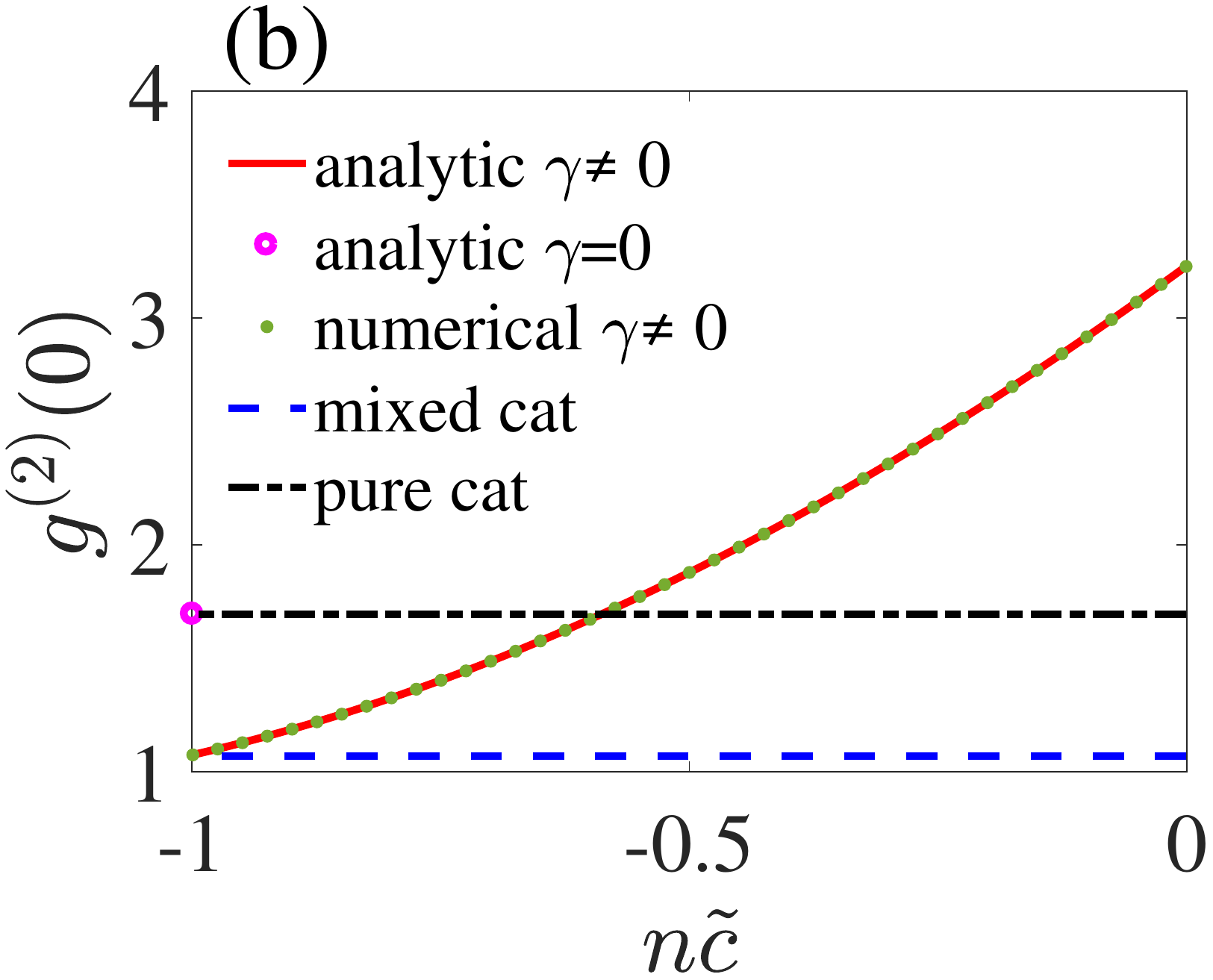}

\includegraphics[width=0.23\textwidth]{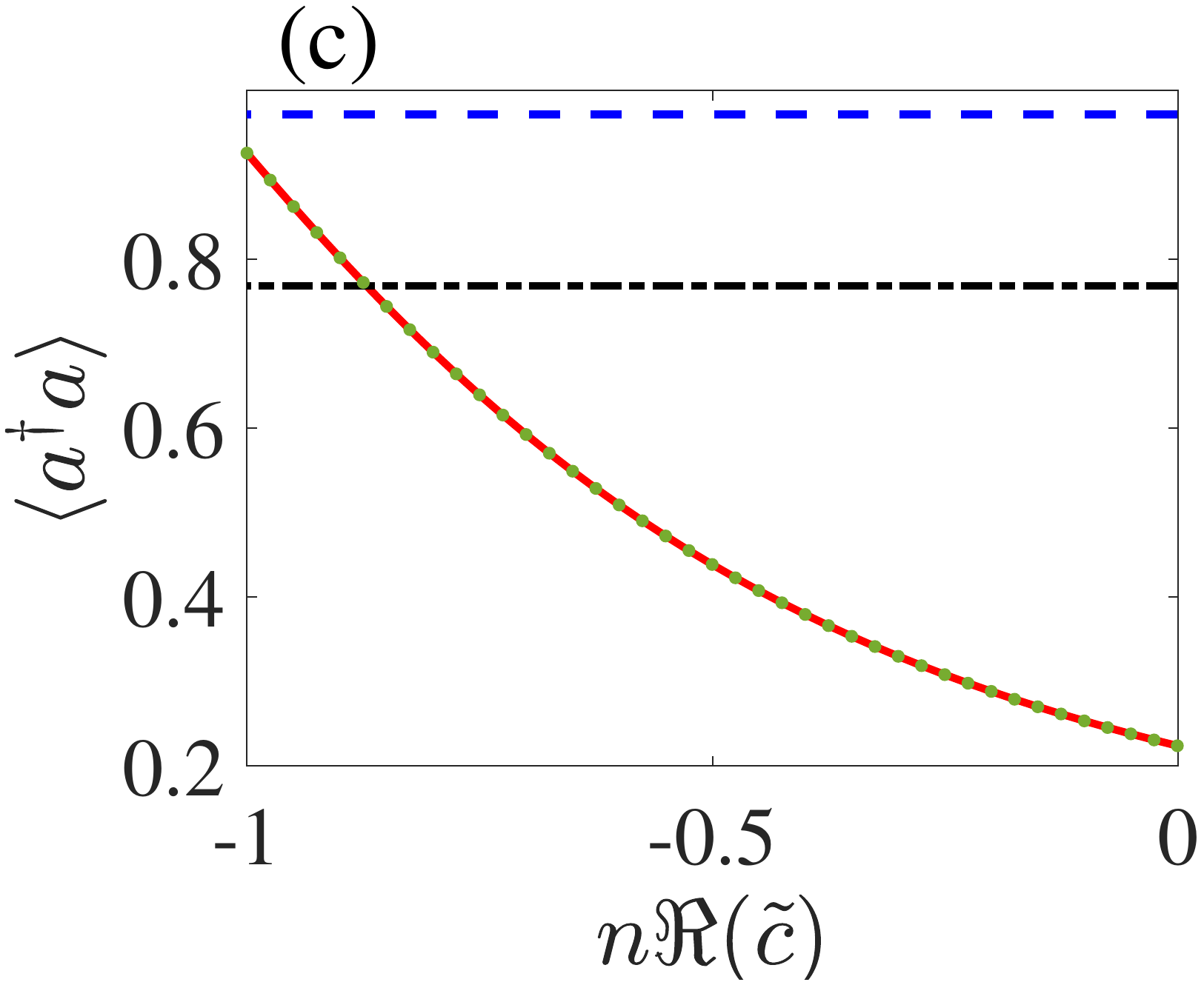} \includegraphics[width=0.23\textwidth]{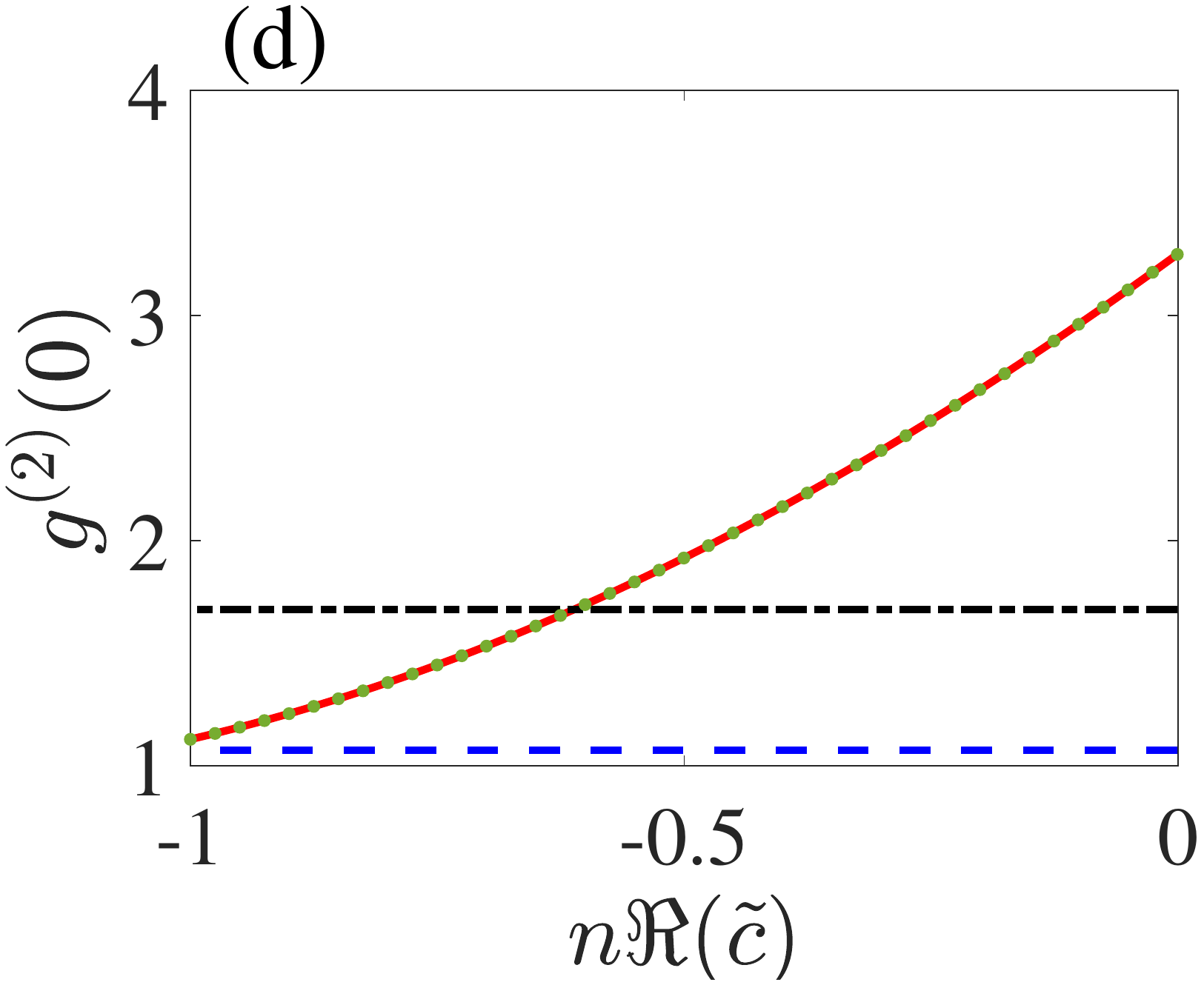}

\includegraphics[width=0.23\textwidth]{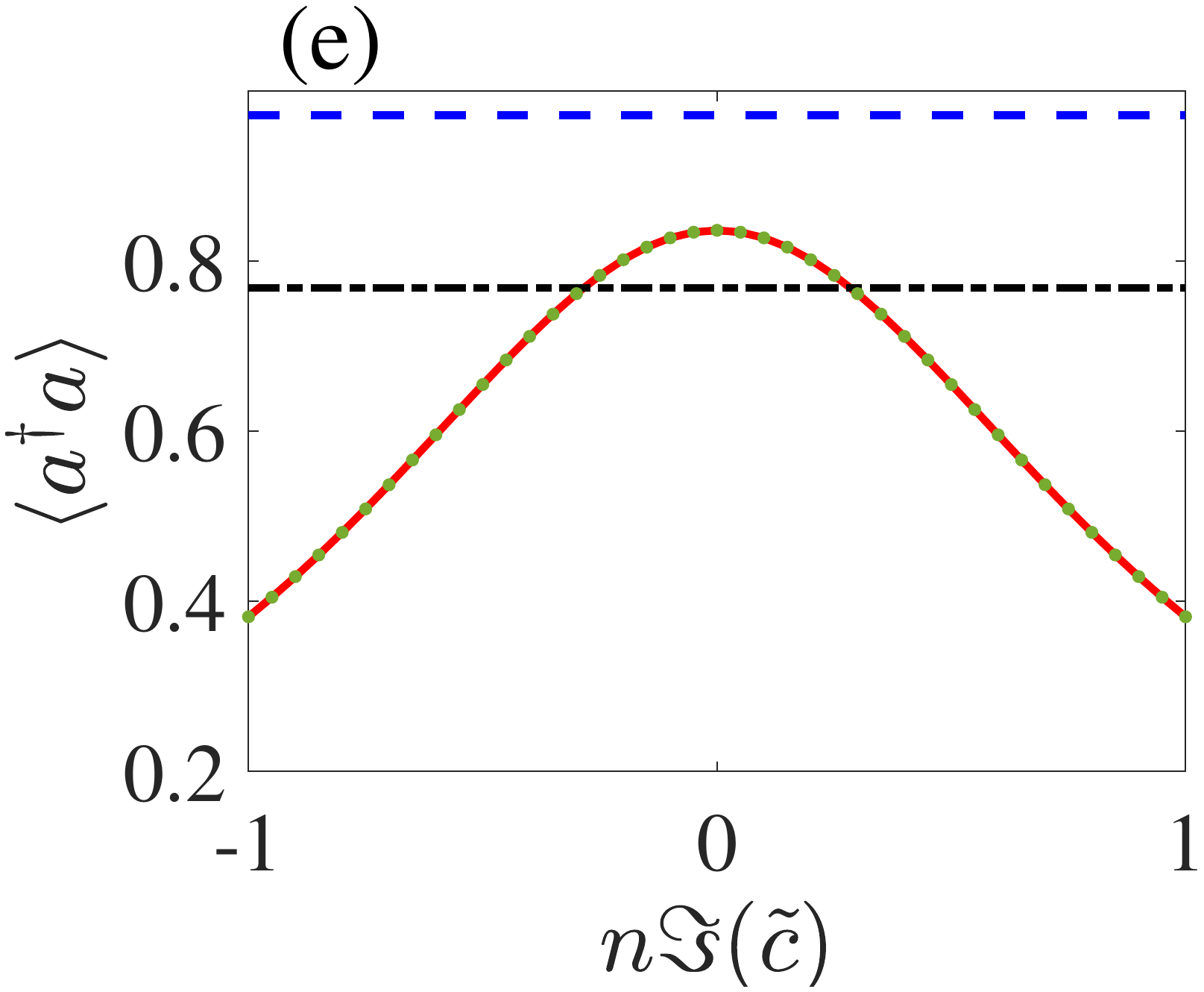} \includegraphics[width=0.23\textwidth]{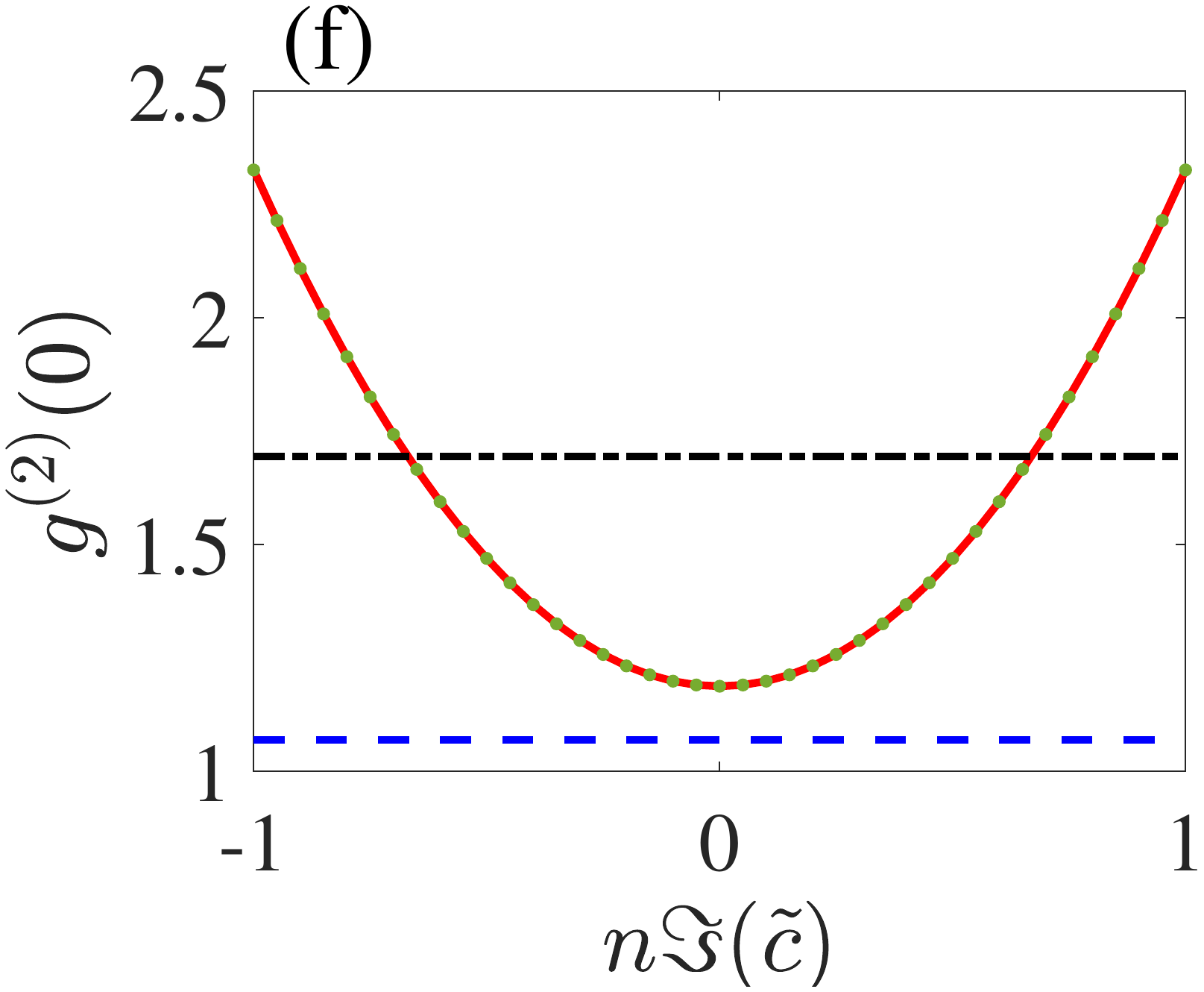}
\caption{Comparisons of the average photon numbers~(a, c, e) and second order
correlation functions~(b, d, f). In figures (a) and (b), the results
are changing with $\tilde{c}$ real. In figures (c) and (d), the real
part of $\tilde{c}$ is changed with $\Im(\tilde{c})=-0.199$. In
figures (e) and (f), the imaginary part of $\tilde{c}$ is changing,
with $\Re(\tilde{c})=-0.896$. The driving $\epsilon=1+0.1i$ in all
figures. Since $n=\left|\epsilon\right|$ is fixed, we scale by$n$
on the x-axis so the limit is simply $n\tilde{c}\to-1$. The blue
dashed line is obtained from the delta-function distribution~(\ref{eq:moment-delta}),
the red solid line from the exact method~(\ref{eq:moment-expand}),
the green dotted line is obtained from the numerical solution, and
the black dash-dotted line from the pure cat state~(\ref{eq:moment-cat}).
The magenta circles in (a) and (b) are obtained from the results~(\ref{eq:distribution_nodecay})
with $\gamma=0$ and an initial vacuum state.}
\label{fig:compare-c}
\end{figure}

We have compared the average steady-state photon number $\left\langle a^{\dagger}a\right\rangle $
and the second order correlation function $g^{(2)}(0)$ changing with
$c$ in Fig.~\ref{fig:compare-c}. The results of Fig.~\ref{fig:compare-c},
show that the delta-function distribution~(\ref{distribution}) is
only attainable when $\tilde{c}\to-1/n$, which is valid when $\gamma_{1}^{(1)}\ll\gamma_{e}^{(2)}$
or $\gamma_{1}^{(1)}\ll\chi_{e}$, if there are no detunings. Mathematically,
it is obtained by reaching the steady state first and then taking
the limit $\gamma_{1}^{(1)}\to0$, which is different from the magenta
circles where we take $\gamma_{1}^{(1)}=0$ exactly and then get the
steady states assuming some particular parity~\citep{gilles1994generation}.
Number parity is conserved only if $\gamma_{1}^{(1)}=0$, and non-conserved
if $\gamma_{1}^{(1)}\neq0$. Thus the ordering of the limit is important,
which leads to the gap between the red line with $\tilde{c}\to-1/n$
(a mixed state) and the magenta circles (a pure cat state) in Fig.~\ref{fig:compare-c}.
In addition, the delta-function distribution can also be obtainable
in the region of extremely strong nonlinearity as the limit $\tilde{c}\to-1/n$
suggests, which is more practical than the case $\gamma_{1}^{(1)}=0$.

In Fig.~\ref{fig:compare-c} the results of the delta-function distributions
never agree with those of the cat states. This is consistent with
the discussion above that the steady state of the system is always
a mixed state~(\ref{eq:mixed-cat}) instead of a pure cat state.
Although there are crosses for the exact results of the steady state
and those of the pure cat state, they are always at different $\tilde{c}$
for $\left\langle a^{\dagger}a\right\rangle $ and $g^{(2)}(0)$.
The exact steady state is therefore different from both the cat state
and a mixture of delta-functions. Hence we can't generate a pure steady-state
cat state, unless the system has no single-photon losses.

\begin{figure}[t]
\centering \includegraphics[width=0.35\textwidth]{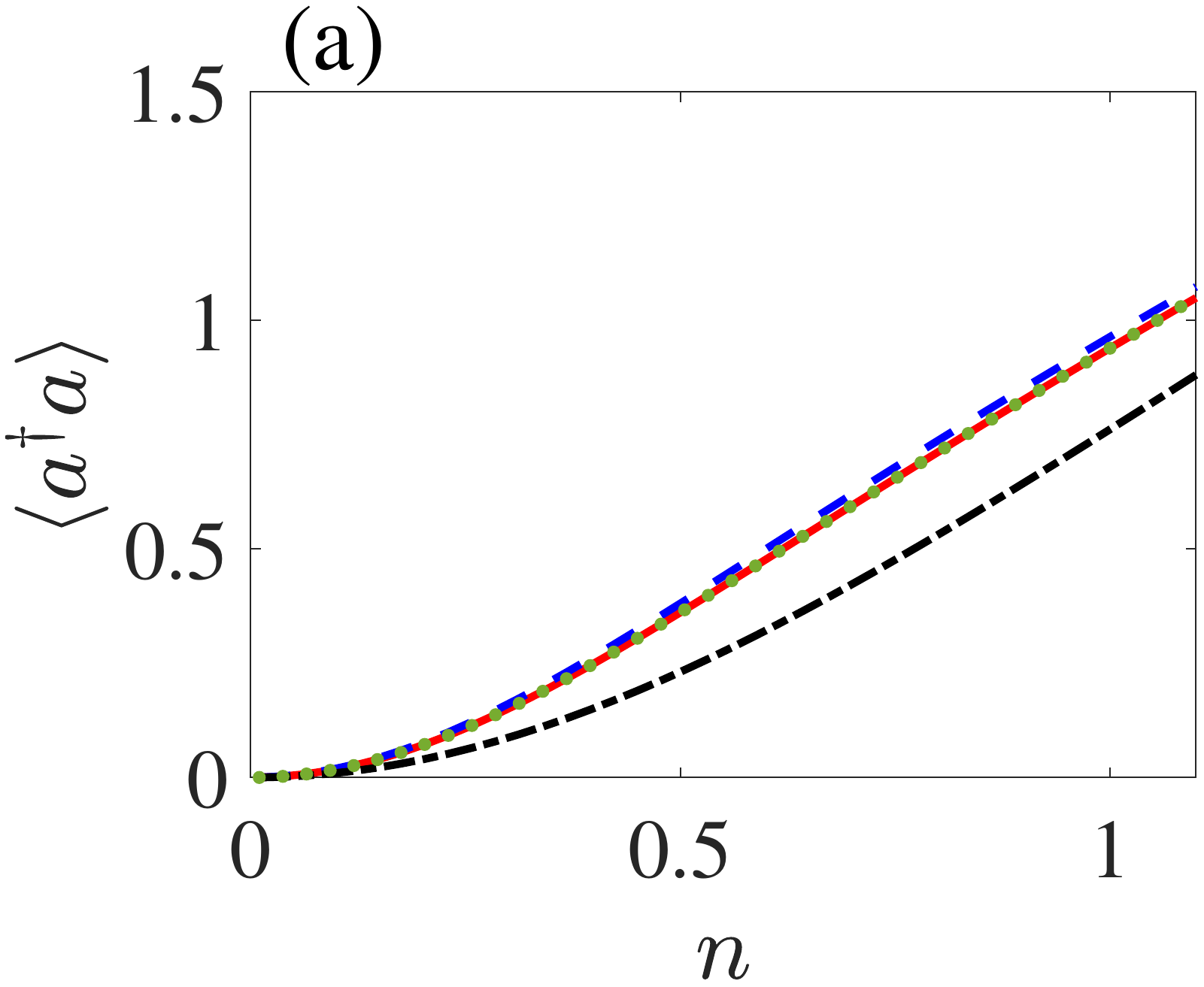}

$\vphantom{}$

\includegraphics[width=0.35\textwidth]{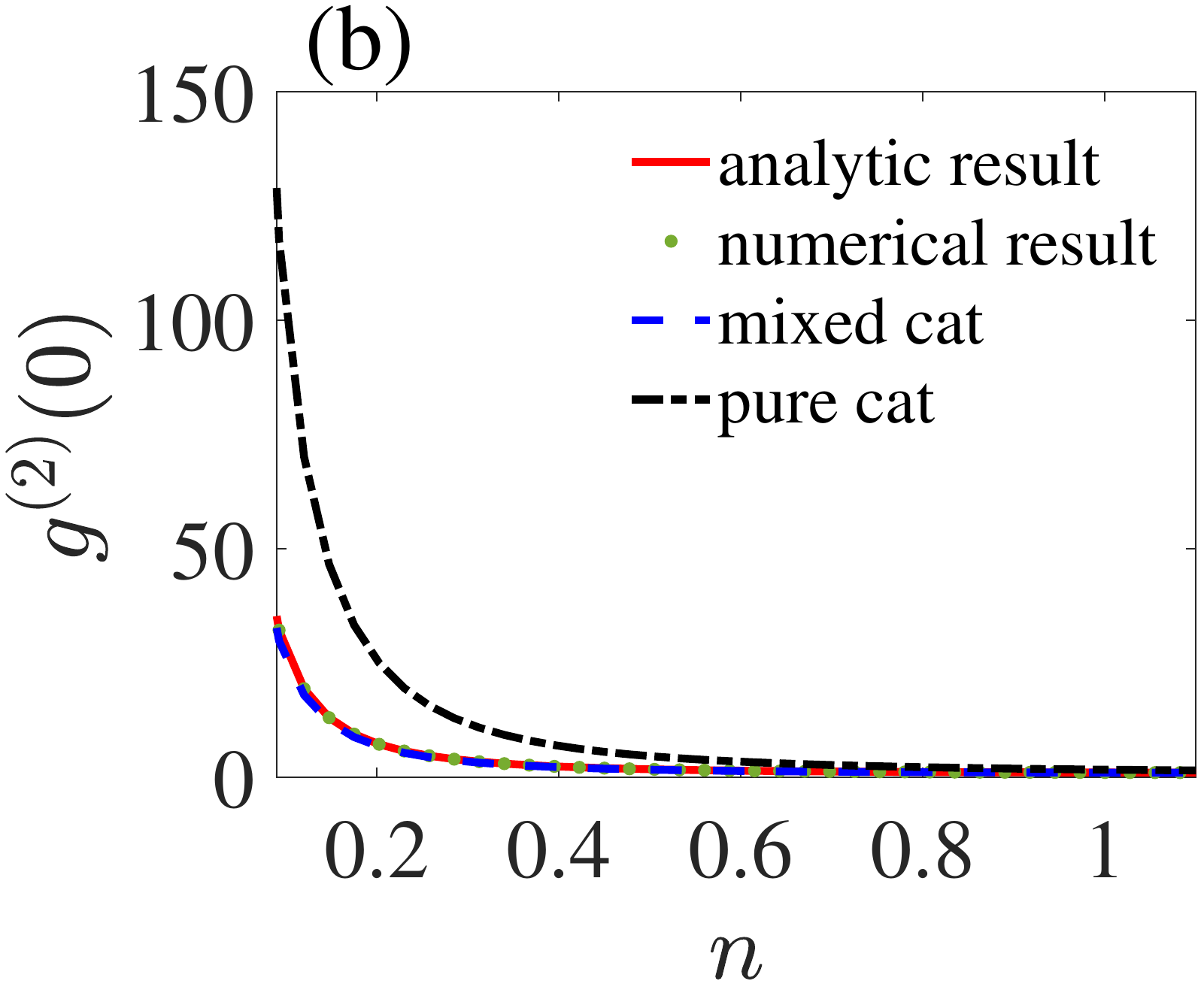} \caption{Comparing the average photon number~(a) and the second-order correlation
function~(b) with $n$ varying. In this case, $n\tilde{c}=-0.99-0.1i$
thus it is close to the limit $n\tilde{c}\to-1$. The lines have the
same meanings as in the Fig.~\ref{fig:compare-c}.}
\label{fig:compare-lambda}
\end{figure}

We have stated that in the limit of small $n$, the delta-function
distribution~(\ref{distribution}) tends to an approximate Schr\" odinger
cat. Now we show how $\left\langle a^{\dagger}a\right\rangle $ and
$g^{(2)}(0)$ change with $n$ in Fig.~\ref{fig:compare-lambda}.
It is natural that the average photon number $\left\langle a^{\dagger}a\right\rangle $
increases with large driving $\mathcal{E}_{2}\propto n$ as shown
in Fig.~\ref{fig:compare-lambda}(a). It also shows that in the region
of small $n$, their photon numbers agree with each other, but $g^{(2)}(0)$
has a different behavior.

This means that even with $n\to0$, the delta-function steady-state
distribution~(\ref{distribution}) is still different from the distribution
of a Schr\" odinger cat. We also show in Fig.~\ref{fig:compare-lambda}
that in the limit of $\tilde{c}\to-1/n$, the exact steady state will
approach the delta-function steady-state distribution, although as
before, this is not a cat state.

It is directly checked with Eqs.~(\ref{eq:moment-delta}) and (\ref{eq:moment-cat})
that the second-order correlation functions are 
\begin{equation}
g_{lim}^{(2)}(0)=\left(\frac{e^{4n}+1}{e^{4n}-1}\right)^{2},\;g_{cat}^{(2)}(0)=\left(\frac{e^{2n}+1}{e^{2n}-1}\right)^{2}.
\end{equation}
Thus in the limit of $n\to0$, we have $g_{cat}^{(2)}(0)/g_{lim}^{(2)}(0)\to4$
with $g_{cat}^{(2)}(0)\to\infty$ and $g_{lim}^{(2)}(0)\to\infty$.
This tendency can be found in the Fig.~\ref{fig:compare-lambda}.
In addition, we will also have $g_{cat}^{(2)}(0)>g_{lim}^{(2)}(0)>1$
over the full range of $n$. This means that their probability distributions
are both super-Poissonian~\citep{scully1997quantum}. From all the
discussions above, we demonstrate that the delta-function steady-state
distribution~(\ref{distribution}) is different from the Schr\" odinger
cat state, even if $n\to0$.

Pure steady-state cats can occur in systems without single-photon
loss and anharmonic nonlinearity~\citep{gilles1994generation}. If
we neglect the single-photon loss in our system from the beginning,
the steady-state solution is obtained from solving $\partial\rho_{1}/\partial t=0$
in Eq.~(\ref{eq:master_eq}). We expand the density operator in the
coherent state basis as $\rho_{1}\left(t=\infty\right)=\iint c_{\alpha,\alpha'}|\alpha\rangle\langle\alpha'|\,d^{2}\alpha d^{2}\alpha'$.
Substituting into Eq. (\ref{eq:master_eq}) with $\gamma_{1}^{(1)}=0$,
for arbitrary $c_{\alpha,\alpha'}$ we have
\begin{align}
\alpha & =\pm\sqrt{\epsilon},\;\alpha'=\pm\sqrt{\epsilon}.
\end{align}
Thus the steady-state density matrix with no single-photon damping
takes the form, 
\begin{eqnarray}
\rho_{1}\left(\infty\right) & = & c_{++}|\sqrt{\epsilon}\rangle\langle\sqrt{\epsilon}|+c_{--}|-\sqrt{\epsilon}\rangle\langle-\sqrt{\epsilon}|\\
 &  & +c_{-+}|-\sqrt{\epsilon}\rangle\langle\sqrt{\epsilon}|+c_{+-}|\sqrt{\epsilon}\rangle\langle-\sqrt{\epsilon}|\,,\nonumber 
\end{eqnarray}
where the coefficients $c_{\alpha,\alpha'}$ are determined by the
initial states. This is consistent with earlier work~\citep{gilles1994generation},
which however had no Kerr anharmonic term. In the P-representation,
the distribution reads in this undamped case, 
\begin{eqnarray}
P_{\infty}(\beta,\beta^{+}) & = & c_{++}\delta(\beta-1)\delta(\beta^{+}-1)\nonumber \\
 &  & +c_{--}\delta(\beta+1)\delta(\beta^{+}+1)\label{eq:distribution_nodecay}\\
 &  & +c_{+-}e^{-2n}\delta(\beta-1)\delta(\beta^{+}+1)\nonumber \\
 &  & +c_{-+}e^{-2n}\delta(\beta+1)\delta(\beta^{+}-1),\nonumber 
\end{eqnarray}
 which is also a delta-function distribution. The possible pure state
solutions are coherent states and cat states. Since the parity is
conserved without single-photon loss according to the master equation~(\ref{eq:master_eq}),
 Schr\" odinger cats can be achieved if the initial states are eigenstates
of the parity, such as Fock states. These steady-state Schr\" odinger
cats with $\gamma_{1}^{(1)}=0$ and initial vacuum states have been
graphed in Fig.~\ref{fig:compare-c} (a) and (b), where a gap between
them and the results for the limit $\gamma_{1}^{(1)}\to0$, which
is a mixture, can be observed.

\section{\label{sec:Summary}Summary}

We have studied the steady states of quantum subharmonic generation
with strong nonlinearity, which has been experimental achieved~\citep{leghtas2015confining}.
By comparing the correlation functions, we conclude that true Schr\" odinger
cats cannot survive in the steady state unless there is no single-photon
loss. With single-photon loss included, the steady state for subharmonic
generation will reduce to a delta-function steady-state distribution~(\ref{distribution})
only if there is an extremely strong nonlinearity. More generally,
the exact solution is always more complex than any type of delta-function,
whether a pure or mixed state. To obtain this exact behavior, the
correct integration manifold is a Pochhammer contour which samples
both sheets of a double Riemann sheet contour. Intriguingly, this
reflects some of the character of the transient macroscopic superposition
that occurs on the path to the steady-state.
\begin{acknowledgments}
This work is supported by the National Key R\&D Program of China (Grants
No. 2016YFA0301302 and No. 2018YFB1107200), the National Natural Science
Foundation of China (Grants No. 11622428, No. 61475006, and No. 61675007)
and the Graduate Academic Exchange Fund of Peking University. PDD
and MDR thank the Australian Research Council and the hospitality
of the Institute for Atomic and Molecular Physics (ITAMP) at Harvard
University (supported by the NSF), and the Weizmann Institute of Science.
This research has also been supported by the Australian Research Council
Discovery Project Grants schemes under Grants DP180102470 and DP190101480.
\end{acknowledgments}

\bibliographystyle{apsrev4-1}
%

\end{document}